\documentclass[11pt,twoside,nohyper]{JHEP1}
\DeclareUnicodeCharacter{0096}{-}
\usepackage{epsfig}
\usepackage{amssymb}
\usepackage{latexsym}
\usepackage{times}
\usepackage{slashed,cite}
\usepackage{amsmath}
\usepackage{amsfonts}
\usepackage{amsbsy}
\usepackage{amscd}
\usepackage{bbm}
\usepackage{graphicx}
\usepackage{epstopdf}
\usepackage{grffile}

\newcommand{\eem}{\end{matrix}}
\newcommand{\bem}{\begin{matrix}}
\newcommand{\eeq}{\end{equation}}
\newcommand{\beq}{\begin{equation}}
\newcommand{\ba}{\begin{array}}
\newcommand{\ea}{\end{array}}
\newcommand{\bea}{\begin{eqnarray}}
\newcommand{\eea}{\end{eqnarray}}
\newcommand{\baq}{\begin{eqnarray}}
\newcommand{\eaq}{\end{eqnarray}}
\newcommand{\beqs}{\begin{subequations}}
\newcommand{\eeqs}{\end{subequations}}

\newcommand{\ecs}{\end{cases}}
\newcommand{\bcs}{\begin{cases}}

\newcommand\eqs[2]{Eqs.~(\ref{#1}) and (\ref{#2})}

\newcommand{\ftn}{\footnotesize}

\newcommand{\TeV}{{\mbox{\rm TeV}}}

\newcommand{\GeV}{{\mbox{\rm GeV}}}
\newcommand{\YeV}{{\mbox{\rm YeV}}}

\newcommand{\etal}{{\it et al.\/}}

\def\to{\rightarrow}

\def\lf{\left(}
\def\rg{\right)}

\newcommand\vev[1]{\langle {#1} \rangle}
\newcommand\vevi[1]{\langle {#1} \rangle_{\rm I}}

\newcommand{\Vhi}{\ensuremath{V_{\rm I}}}
\newcommand{\vf}{\ensuremath{V_{\rm F}}}
\newcommand{\vfo}{\ensuremath{V_{\rm F0}}}

\newcommand{\dq}{\ensuremath{d_{q}}}

\newcommand{\ck}{\ensuremath{c_{2K}}}

\newcommand{\ns}{\ensuremath{n_{\rm s}}}
\newcommand{\ks}{\ensuremath{k_\star}}

\newcommand{\bph}{\ensuremath{\bar \Phi}}

\newcommand{\tra}{\ensuremath{{\sf T}^{\rm a}}}

\newcommand{\phc}{\ensuremath{\Phi}}
\newcommand{\psc}{\ensuremath{\Psi}}
\newcommand{\phcb}{\ensuremath{\bar \Phi}}
\newcommand{\Trh}{\ensuremath{T_{\rm rh}}}

\newcommand{\sg}{\ensuremath{\sigma}}
\newcommand{\sgx}{\ensuremath{\sigma_\star}}

\newcommand{\sgf}{\ensuremath{\sigma_{\rm f}}}

\def\K{{\hat{K}}}

\def\Z{\hat{Z}}

\def\mP{m_{\rm P}}

\newcommand{\bJ}{\ensuremath{\bar J}}

\newcommand{\Eref}[1]{Eq.~(\ref{#1})}
\newcommand{\Sref}[1]{Sec.~\ref{#1}}
\newcommand{\Fref}[1]{Fig.~\ref{#1}}
\newcommand{\Tref}[1]{Table~\ref{#1}}
\newcommand{\cref}[1]{Ref.~\cite{#1}}

\def\fhi{smFHI}
\def\fhia{smFHI-I}
\def\fhib{smFHI-II}
\def\shsgr{shSUGRA}
\def\nsgr{NSUGRA}
\def\msgr{mSUGRA}

\def\Kas{K\"{a}hler potentials}

\def\Ka{K\"{a}hler potential~}
\def\Kaa{K\"{a}hler}
\def\Kam{K\"{a}hler manifold}
\def\Kap{K\"{a}hler potential}

\def\actc{{\sf\ftn P-ACT-LB-BK18}}
\def\spt{{\sf\ftn P-ACT-SPT}}
\def\actb{{\sf\ftn P-ACT-LB}}

\newcommand{\plk}{{\it Planck}}

\renewcommand{\nc}{\ensuremath{N_{0}}}
\newcommand{\as}{\ensuremath{a_{\rm s}}}
\newcommand{\As}{\ensuremath{A_{\rm s}}}
\newcommand{\Tr}{\mbox{\sf Tr}}

\newcommand{\Vhio}{\ensuremath{V_{\rm I0}}}
\newcommand{\Hhi}{\ensuremath{H_{\rm I0}}}
\newcommand{\ckf}{\ensuremath{c_{4K}}}
\newcommand{\ckx}{\ensuremath{c_{6K}}}
\newcommand{\ckh}{\ensuremath{c_{8K}}}
\newcommand{\kp}{\ensuremath{\kappa}}

\newcommand{\mst}{\ensuremath{M_*}}

\newcommand{\mstr}{\ensuremath{M_{\rm S}}}
\newcommand{\msn}{\ensuremath{m_{\rm I}}}
\newcommand{\sgb}{\ensuremath{\bar \sigma}}

\newcommand{\cpa}{\ensuremath{c_{1p}}}
\newcommand{\cpb}{\ensuremath{c_{2p}}}
\newcommand{\cpd}{\ensuremath{c_{4p}}}

\newcommand{\cqa}{\ensuremath{c_{1q}}}
\newcommand{\cqb}{\ensuremath{c_{2q}}}
\newcommand{\cqd}{\ensuremath{c_{4q}}}

\newcommand{\ep}{\ensuremath{\epsilon}}

\newcommand{\Nhi}{\ensuremath{N_{\rm I\star}}}
\newcommand{\Ns}{\ensuremath{N_{\star}}}
\newcommand{\al}{\ensuremath{\alpha}}
\newcommand{\bt}{\ensuremath{\beta}}
\newcommand{\ca}{\ensuremath{{\rm a}}}

\newcommand{\Ggut}{\ensuremath{\mathbb{G}}}

\newcommand{\sun}{\ensuremath{SU({\cal N})}}

\newcommand{\diag}{\ensuremath{{\sf diag}}}
\newcommand{\wrh}{\ensuremath{w_{\rm rh}}}

\title{GUT-Scale Smooth Hybrid Inflation with a Stabilized Modulus in Light of ACT and SPT Data}

\author{Waqas Ahmed,$^{\sf(1)}$ Constantinos Pallis,$^{\sf(2)}$ and Mansoor Ur Rehman$^{\sf(3)}$ \\
$^{\sf(1)}$ Center for Fundamental Physics and School of Mathematics and Physics, \\
Hubei Polytechnic University, Huangshi 435003, China; \\ E-mail: \email{waqasmit@hbpu.edu.cn}   \\
$^{\sf(2)}$ School of Technology,  Aristotle University of
Thessaloniki, \\ Thessaloniki, GR-54124 Greece;
\\ E-mail: \email{kpallis@auth.gr}\\
$^{\sf(3)}$ Department of Physics, Faculty of Science, Islamic
University of Madinah, \\ Madinah 42351, Saudi Arabia; \\ E-mail:
\email{m.rehman@iu.edu.sa}}

\abstract{We analyze a generalized framework of smooth F-term
hybrid inflation (smFHI) consistent with gauge coupling
unification within the Minimal Supersymmetric Standard Model
(MSSM). The embedding of the model in two specific
Supergravity settings addresses at the same time the
$\eta$ problem and the compatibility with the recent ACT or SPT data.
The one relies on the choice of a shift-symmetric \Ka for the
inflaton which revitalizes the SUSY predictions of smFHI, whereas
the other employs a \Ka associated with an hyperbolic \Kaa\ manifold.
An essential role in both our constructions is played by a
decoupled superheavy field without superpotential and \Ka inspired
by string- and D-brane-based models. Our proposal can be realized
for a variety of representations for the Higgs fields involved in
\fhi\ and assures monotonic inflationary potential.\\\\

{\sc Published in:} {{\sl  J. Cosmol. Astropart. Phys.} {\bf 06},
048 (2026)}

}
\begin{document}

\setcounter{page}{1} 



\section{Introduction}\label{intro}

Among the various versions \cite{lect,hinova} of \emph{F-term
hybrid inflation} ({\ftn\sf  FHI}) the so-called smooth
\cite{sm,sm1,ini,nsm,
smy,smk,smii,smsu5b,senoguz,simple,okada,mur,simple5,zubair}
is well distinguishable and observationally
more promising -- for a variant called pseudosmooth tribrid
inflation see \cref{pseudo1,pseudo2}. This is because the
inflationary path displays a classical inclination which drives
the inflaton $S$ towards its vacuum smoothly -- contrary to the
standard \cite{fhi} and shifted \cite{sh, nsh, ssh, su4sh, su5sh,
mush} versions of FHI which are tied to a classically flat
trajectory which develops an instability at some point terminating
abruptly FHI. As a consequence, the (scalar) spectral index,
$n_{\rm s}$ within \emph{smooth FHI} ({\ftn\sf smFHI}) in the
context of \emph{supersymmetry} ({\ftn\sf SUSY}) turns out to be
significantly lower than the other two models. Namely, we obtain
\cite{sm,hinova}
\beq \label{nssm} \ns\simeq1-5/3\Ns=0.967\>\>\mbox{with}\>\>\Ns=50
\eeq
being the number of e-foldings that the pivot scale $\ks$ suffered
during \fhi. Moreover, \fhi\ offers a parametric flexibility so
that the \emph{vacuum expectation values} ({\sf\ftn v.e.vs}) of
the inflaton-accompanying Higgs fields are less restricted by the
normalization of the power spectrum of the curvature perturbation
compared to the situation in the standard FHI \cite{hinova}. As a
result, these v.e.vs can become \cite{sm,hinova} precisely equal
to the value suggested by the unification of the gauge coupling
constants within SUSY \emph{Grand Unified Theories} ({\ftn\sf
GUTs}) adopting the particle content of the \emph{Minimal SUSY
Standard Model} ({\ftn\sf MSSM}) up to the unification scale.

The result in \Eref{nssm} is tantalizing close to the central
value reported by the \plk\ data \cite{plin} combined with
\emph{BICEP/Keck} ({\sf\ftn BK}) \cite{bcp} according to which
\beq \label{nspl} \ns\simeq0.9652\pm0.0084 ~~\mbox{at 95$\%$
\emph{confidence level} ({\sf\small c.l.})}.\eeq
This value, however, is substantially lower than the so-called
\actc\ data which includes the latest \emph{Data Release 6}
({\sf\ftn DR6}) from the \emph{Atacama Cosmology Telescope}
({\sf\ftn ACT}) \cite{act,actin}, combined with the \emph{cosmic
microwave background} ({\sf\ftn CMB}) measurements by \plk\
\cite{plin} and \emph{BICEP/Keck} ({\sf\ftn BK}) \cite{bcp},
together with the \emph{Dark Energy Spectroscopic Instrument}
({\sf\ftn DESI}) {\it Baryon Acoustic Oscillation} ({\sf\ftn BAO})
results \cite{desi}. Namely, the \actc\ data entails \cite{actin}
\beq \label{nsact}
\ns=0.9743\pm0.0068\>\Rightarrow\>0.967\lesssim\ns\lesssim0.981\>\>\mbox{at
95$\%$ c.l.} \eeq
On the other hand, the \emph{South Pole Telescope} ({\sf\ftn SPT})
announced recently \cite{spt} new measurements. Combining these
with the ones from \plk\ and ACT, we obtain the \spt\ data which
dictates
\beq \label{nsspt}
\ns=0.9684\pm0.006\>\Rightarrow\>0.962\lesssim\ns\lesssim0.974\>\>\mbox{at
95$\%$ c.l.} \eeq
A wide stream of works \cite{act0,
act2,act4,gup,oxf,indi,kina,actlee,rhc,rhb,
rha,act5,nmact,maity,reh8, actattr,act1,actj,yin,actpal,
act3,act6,act7,act8,actellis,acttamv,ketov,
r2a,r2b,r2drees,r2mans,r2li,heavy,fhi1,fhi2,fhi3,fhi4,fhi5,actpole,phi,das,
salvio,aoki,hai} appeared recently trying to reconcile several
well-motivated inflationary models with the observational limits
above -- mainly that of \Eref{nsact}.

Aspiring to establish realistic versions of \fhi, we have to
remind that the inclusion of \emph{supergravity} ({\ftn\sf SUGRA})
corrections with canonical K\"ahler potentials -- within the
so-called \emph{minimal SUGRA} ({\ftn\sf mSUGRA}) scenario --
typically pushes $n_{\rm s}$ above the aforementioned
observational margins. This occurs despite the fact that the
generation of a mass term for $S$ is avoided thanks to a mutual
cancellation. This upward shift in $n_{\rm s}$ can, however, be
circumvented if the analysis is restricted to scales well below
the SUSY GUT scale. In this regime \cite{okada} the \actc\ results
in \Eref{nsact} can be marginally reproduced for $\Ns=55$.
Alternatively, employing a quasi-canonical \Ka \cite{senoguz, nsm,
simple, mur} with a suitable choice of the sign of the
next-to-minimal terms allows \fhi\ to agree with observational
data above. This setup, however, leads to hilltop-type
\cite{lofti} solutions over most of the parameter space
introducing thereby, an inherent fine-tuning of the initial
conditions, related to the maximum -- minimum structure of the
inflationary potential -- cf. \cref{hinova,nsm}. Insisting on the
one step inflationary paradigm -- cf. \cref{fhim} -- a more drastic,
and arguably more radical, approach to circumvent the $\ns$
problem of \fhi\ is the inclusion of additional fields into the
model which exhibit no-scale-type \Kas. These fields naturally
arise as moduli in string theory \cite{ibanez,lust} and can be
assumed stabilized \cite{kelar,hinova,antu, eno7} during \fhi.
They contribute extra terms to the SUGRA corrections of the
inflationary potential \cite{hpana1,hpana2, asfhi, hinova, kelar}
affecting thereby the inflationary observables. An added advantage
of this construction is the possibility of evading the notorious
$\eta$ problem \cite{eta} of FHI by constraining the coefficients
of $K$ to natural values (of order unity).

Working along the lines of the latter idea, we here present novel
variants of \fhi\ which may be reconciled with the data in
\Eref{nsspt} or \Eref{nsact}. In the first variant, the inclusion
of one modulus together with a shift-symmetric $K$ \cite{shiftk0,
shiftk} for the inflaton assist to identify the $\ns$ value with
that in \Eref{nssm}. We call this setting \emph{shift-symmetric
SUGRA} ({\ftn\sf shSUGRA}). In the second variant, the utilization
of an hyperbolic \Ka \cite{hpana1,nsh, eno7,alinde} for the
inflaton assures a slight positive shift of $\ns$ without
resorting to hilltop solutions. This scenario is named
\emph{$N$-depended SUGRA} ({\ftn\sf NSUGRA}). Our framework also
enables us to extend \fhi\ to alternative representations of the
relevant Higgs fields -- cf. \cite{sm1, smii, smsu5b, simple5} --
and to explore various choices of the associated exponents -- cf.
\cref{smy, okada}. Furthermore, we verify the viability of \fhi\
with a cutoff scale in the superpotential as high as the Planck
mass, $\mP$. A notable byproduct of our setting is that the v.e.vs
of the Higgs fields can be fixed exactly at the SUSY GUT scale.
Imposing this condition significantly reduces the parameter space
of the model, thereby enhancing its predictive power.

In what follows, we present the salient features of our set-up in
\Sref{set} and derive the corresponding inflationary potential
(Sec.~\ref{sugra}). We then outline the observational constraints
relevant to our setup (Sec.~\ref{cons}), before proceeding to our
numerical analysis (Sec.~\ref{res}) and concluding remarks
(Sec.~\ref{con}). In Appendices \ref{app}, \ref{msgr} and
\ref{lepto} we propose a sample stabilization of the modulus
before \fhi, we clarify the status of GUT-scale \fhi\ within
\msgr\ and we display a representative realization of reheating.

\section{Set-up}\label{set}

We here present the building blocks of our set-up. Namely, in
\Sref{set1} we introduce the superpotentials for the two types of
\fhi\ whereas in \Sref{set2} we quote the K\"ahler potentials
which define the two SUGRA versions, \shsgr\ and \nsgr.

\subsection{Superpotentials}\label{set1}

Smooth FHI can be realized by adopting one of the following
superpotentials \cite{sm, smy, okada, simple5, smsu5b}:
\begin{equation} \label{Whi} W = \bcs
S\left(\mst^{2(1-p)}(\bph\Phi)^p-M^2\right)&\mbox{for Type I \fhi\ ({\sf\ftn \fhia}),} \\
S\left(\mst^{(2-q)}\Tr(\Psi)^q - M^2\right)&\mbox{for Type II
\fhi\ ({\sf\ftn \fhib}).} \ecs
\end{equation}
Here $p$ and $q$ are integers with $p>1$ and $q>2$. In case of
\fhia, the fields $\bar{\Phi}$ and $\Phi$ form a pair of
left-handed superfields belonging to non-trivial conjugate
representations of the GUT gauge group $\Ggut$, reducing its rank
through their v.e.vs -- see e.g. \cref{sm1, nsm, zubair}. By
contrast, in \fhib\ the Higgs superfield $\Psi=\Psi^\ca \tra$
transforms in the adjoint representation of $\Ggut$, which may be
identified with e.g. $SU(2)_{\rm  R}$ of $SU(3)_{\rm c} \times
SU(2)_{\rm L} \times SU(2)_{\rm  R} \times U(1)_{B-L}$ \cite{smii}
or $SU(5)$ \cite{su5sh, smsu5b, simple5}, or $SU(4)_{\rm c}$ of
$SU(4)_{\rm c} \times SU(2)_{\rm  L} \times SU(2)_{\rm R}$ or
$SU(4)_{\rm c} \times SU(2)_{\rm L} \times U(1)_{\rm R}$
\cite{su4sh}.  For the adjoint representation, we adopt the
normalization
\beq  \Tr(\tra {\sf T}_{\rm }^{\rm b})=\delta^{\rm
ab}/2~~\mbox{where}~~\ca=1,...,{\cal N}^2-1. \eeq
Here $\Tr$ denotes the trace of the Hermitian matrix $\tra$, and
$\ca$ runs over the dimensionality of the corresponding algebra.
Consequently, the kinetic term of $\Psi$ takes the form
$\Tr(\dot\Psi)^2=(\dot\Psi^\ca)^2/2$, where the dot indicates
derivative with respect to cosmic time.

The field $S$ is a $\Ggut$-singlet left-handed superfield, and
$\mst$ represents an effective cutoff scale, naturally comparable
to the string scale $\mstr=5\cdot10^{17}~\GeV$. We also explore a
simplified version of \fhi\ \cite{simple,simple5}, where
\beq\label{kp} \mst = \bcs \kp^{1/2(1-p)}\mP &\mbox{for \fhia}\\
 \kp^{1/(2-q)}\mP&\mbox{for \fhib}.\ecs \eeq
To ensure the perturbativity we impose on $\kp$ the condition $\kp
< \sqrt{4\pi}$.

The superpotential in Eq.~(\ref{Whi}) is consistent with a
continuous $R$ symmetry \cite{fhi}, under which $S$ and $W$ carry
the same charge, while $(\bar{\Phi}\Phi)^p$ or $\Psi^q$ remain
neutral. In addition, we impose an extra global $\mathbb{Z}_p$ or
$\mathbb{Z}_q$ discrete symmetry, under which $\Phi \rightarrow
e^{2 \pi i/p} \Phi$ or $\Psi\to e^{2 \pi i/q} \Psi$ for \fhia\ or
\fhib, respectively. As a result, only the $p$ or $q$ powers of
$\bar{\Phi}\Phi$ or $\Psi$ correspondingly are allowed.

\subsection{\Kas}\label{set2}

The full \Kas\ of our models include separated contributions for
the inflaton, $K_{\rm I}$, the stabilized modulus, $\K$, and the
Higgs fields $K_{\rm H}$. It can be written as
\beq K=K_{\rm I} +\K+  K_{\rm H}\>\>\mbox{with}\>\>K_{\rm H}=
 \bcs |\phc|^2+|\phcb|^2 &\mbox{for \fhia}\\
\Tr|\Psi|^2 &\mbox{for \fhib.}\ecs\label{ktot}\eeq
The form of $K_{\rm I}$ depends on the SUGRA scenario. We consider
two specific choices which include just quadratic terms as regards
$S$. Namely,
\begin{itemize}

\item For shSUGRA $K_{\rm I}$ assumes the following form
\beq K_{\rm I}=-\frac12\Z\lf S-S^*\rg^2, \label{ksh} \eeq
which breaks the $R$ symmetry of $W$. However, it is invariant
under the following symmetries: $S \to  S + c$ and $S \to -S$,
where $c$ is a real number.

\item For NSUGRA $K_{\rm I}$ assumes the following form
\beq K_{\rm I} = N \mP^2 \ln\left(1 + {\Z |S|^2}/{N \mP^2}\right),
\label{kn} \eeq
which respects the $R$ symmetry of $W$ and parameterizes either a
compact ($N>0$) or non-compact (hyperbolic, $N<0$) moduli \Kam.
\end{itemize}
The functions $\K$ in \Eref{ktot} and $\Z$ in \eqs{ksh}{kn} above
encode the dependence of $K$ on the stabilized modulus $h$. We
assume that $h$ is charged under some extra symmetry and so it is
not present in $W$ \cite{hpana1,hpana2}. Motivated by superstring
and D-brane models \cite{ibanez,lust}, we adopt the ansatz
\beq \label{kh}
\Z=\left({f(h)+f^*(h^*)}\right)^\al\>\>\mbox{and}\>\>\K=\beta\mP^2
\ln\lf f(h)+f^*(h^*)\rg\>\>\mbox{with}\>\>\bt<0,\eeq
where $f$ is an analytic function of $h$, $\beta$ is taken to be
an integer and $\alpha$ a rational number. The restriction $\beta
< 0$ ensures positivity of the kinetic terms of $h$. $\K$
parameterizes the \Kam\ $SU(1, 1)/U(1)$ with constant scalar
curvature $2/\bt$. We assume that $h$ is stabilized during \fhi\
through some mechanism -- see Appendix~\ref{app} for an example.
In particular, we set
\beq \vevi{f(h)}=\vevi{f^*(h^*)}=1/2~~\mbox{so
as}~~\vevi{\Z}=1~~\mbox{and}~~\vevi{\K}=0, \label{vevh}\eeq
where the symbol $\vevi{Q}$ means the value of the quantity $Q$
during \fhi.

Taking into account the form of the various $K$'s above, we can
define the canonically normalized real fields $\sg, \sgb, \phi$
and $\psi$ as follows
\beq \label{cann} S = (\sg+i\sgb)/\sqrt{2}~~\mbox{and}~~ \bcs \phc=\phcb=\phi/2&\mbox{for \fhia,}\\
\Psi^1=\psi&\mbox{for \fhib,}\ecs \eeq
where the MSSM-neutral component of $\Psi$ is aligned along the
$\ca=1$ direction. The imaginary part of $S$ can be absorbed
through an $R$-transformation in case of \nsgr. However,
$R$-symmetry is explicitly broken within \shsgr\ -- see \Eref{ksh}
-- and so the stability of the direction $\sgb=0$ has to be
checked as we see in \Sref{sugra1} below. Since $h$ is stabilized
during \fhi\ there is no $h-S$ mixing within \nsgr\ \cite{kelar,
eno7, hinova}.


\section{Inflationary Potential} \label{sugra}

We here introduce the SUGRA potential of our models in
\Sref{sugra1} and pay special attention in computing the SUSY
contribution in \Sref{susy}. Finally, we obtain a compact
expression for the potential in \Sref{fn} which supports the
various versions of \fhi\ discussed in our work.

\subsection{SUGRA Potential}\label{sugra1}

The F-term SUGRA scalar potential is given by
\begin{equation}
V_{\rm SUGRA}=e^{K/\mP^2}\left(K^{I\bJ}{\rm F}_I {\rm F}^*_{\bJ}
-3\vert W\vert^2/\mP^2\right)+\frac{g^2}2 \mbox{$\sum_{\rm a}$}
{\rm D}_{\rm a} {\rm D}_{\rm a} \label{Vsugra}
\end{equation}
where the subscript $I~[\bJ]$ denotes derivation \emph{with
respect to} ({\ftn\sf w.r.t}) the complex scalar field
$X_I~[X_{\bJ}^{*}]$. Note that we use the same symbol $X_I$ for
the scalar fields and the corresponding chiral superfield, with
$X_I=S,h$ and $\phc,\phcb$ or $\Psi$. The matrix $K^{I\bJ}$ is the
inverse of the K\"ahler metric $K_{I\bJ}=K_{,X^I X^{\bJ}}$ -- with
the symbol $,X^J$ as subscript denoting derivation w.r.t  $X^J$
--, i.e., $K^{I\bJ}K_{I\bar L}=\delta^{\bJ}_{\bar L}$. Also the F
and D terms are
\beq \label{fdsgr} {\rm F}_I =W_{,X^I}
+K_{,X^I}W/\mP^2\>\>\mbox{and}\>\>{\rm D}_{\rm a}=X^I\lf {\sf
T}_{\rm a}\rg^J_I K_J, \eeq
with  $K_{J}={K_{,X^J}}$. Here, the index $I$ should not to be
confused with the subscript ``I'', which is used to denote
evaluation during \fhi.

\fhi\ takes place along a D-flat inflationary trajectory defined
by the conditions
\beq \bcs \vevi{|\phc|-|\phcb|}=0&\mbox{for \fhia,}\\
\vevi{|\Psi^\dagger\Tr[\Psi,\tra]|}^2=0&\mbox{for \fhib,}
\ecs\label{dflat} \eeq
which remain unaffected by the SUGRA corrections, as we consider
canonical kinetic terms for the Higgs fields in \Eref{ktot}. For
\fhib\ the condition is automatically satisfied, owing to the
antisymmetric nature of the \sun\ structure constants and the fact
that all component fields which are not singlets under the
Standard Model gauge group vanish during \fhi.

Taking into account that $W_{,h}=0$, according to our set-up in
\Sref{set1}, and that $\phc, \phcb\ll S$ for \fhia\ and $\Psi\ll
S$ for \fhib, $\vf$, in \Eref{Vsugra} can be written as
\bea \vf \simeq \vfo&+&K^{SS^*}\lf |WK_S/\mP^2|^2+
W_{,S}W^*K_{S^*}/\mP^2+{\rm c.c}\rg\nonumber\\
&+&K^{Sh^*}WK_h/\mP^2\lf W_{,S^*}+W^*K_{S^*}/\mP^2\rg\nonumber\\
&+&K^{hS^*}WK_{h^*}/\mP^2\lf W_{,S}+WK_{S}/\mP^2\rg\nonumber\\
&+&K^{hh^*}|W|^2K_{h}K_{h^*}/\mP^4-3|W|^2/\mP^2,\label{vf}\eea
where $\vfo$ is the SUSY part of $V_{\rm F}$ which is given in
\Sref{susy}. The remaining part of $\vf$ dependents on the SUGRA
scenario. Focusing on the $(S,h)$ subspace of our field space we
below derive $\vevi{K_{I\bJ}}$ and $\vevi{\vf-\vfo}$. Namely,

\begin{itemize}

\item For \shsgr\ we find
\beq \vevi{K_{I\bJ}}=\diag(1,-\bt
\vevi{|f_{,h}|}^2)~~\mbox{and}~~\vevi{\vf-\vfo}=-(3+\bt)M^4
\sg^2/2\mP^2, \label{kvfsh}\eeq
where we take into account $\vevi{K_S}=\vevi{K_{Sh}}=0$ and
$\vevi{K_{SS}}=1$. Note that $\vevi{\vf-\vfo}$ is independent of
$\vevi{f_{,h}}$ although this is present in $\vevi{K_{I\bJ}}$. It
also exhibits a runaway behavior for $\bt=0$ as first noticed in
\cref{shiftk}. From the complete version of $\vf$ we can also
derive the effective mass of $\sgb$ in \Eref{cann} along the
direction $\vevi{\sgb}=0$. We find
\beq m^2_{\sgb}= (M^4/\mP^2) \lf 3 -(2\al-\bt)^2/\bt + ((3 +
\al)\al- \bt-1)(\sg/\mP)^2\rg, \label{m2sgb} \eeq
which depends on $\al$ contrary to $\vevi{\vf-\vfo}$ in
\Eref{kvfsh}.

\item For \nsgr\ we find the following elements of the \Kaa\
metric
\beq\begin{aligned} &\vevi{K_{SS^*}}=\lf1+\sg^2/2 N\mP^2\rg^{-2}\simeq1,\\
&\vevi{K_{Sh^*}}=\vevi{K_{hS^*}}=
(\sg/\sqrt{2})\al\vevi{f_{,h}}\lf1+\sg^2/2 N\mP^2\rg^{-2}\ll1,
\\ &\vevi{K_{hh^*}}=\lf \al\sg^2
\lf2 (\al-1) N- {\sg}^2/{\mP^2}\rg/4N^2\lf1+\sg^2/2 N\mP^2\rg^{2}-\bt\mP^2\rg|\vevi{f_{,h}}|^2\\
&\hspace{1.3cm}\simeq-\bt|\vevi{f_{,h}}|^2\mP^2. \label{vksh}
\end{aligned}\eeq
On the other hand, $\vevi{\vf-\vfo}$ takes the form
\bea \nonumber \vevi{\vf-\vfo}&=& M^4\lf 1 + \frac{\sg^2}{2
N\mP^2}\rg^N \lf
8\bt  N^3- 4 N^2 ((\bt-\al)(1 -\al +\bt)N-3\bt) (\sg/\mP)^2\right. \\
&+& 2 N \lf\bt (3 + N + 2 \al  N + N^2) - \al  N (N+ \al-2 )-\bt^2
N \rg(\sg/\mP)^4 \nonumber \\
&+& \left.(1 + N)^2 (\bt  + \al N) (\sg/\mP)^6\rg/4N^2(2\bt  N +
(\bt + \al  N) (\sg/\mP)^2), \label{vfn} \eea
which is still independent from $\vevi{f_{,h}}$ similarly to our
result in \Eref{kvfsh}.

\end{itemize}

\subsection{SUSY Contribution}\label{susy}

The SUSY F-term scalar potential, $\vfo$, which appears in
\Eref{vf}, takes the form
\beq \label{vsusy} V_{\rm F0}=K^{SS^*}|W_{,S}|^2+\bcs
K^{\Phi\Phi^*}|W_{,\Phi}|^2+K^{\phcb\phcb^*}|W_{,\phcb}|^2 ~~&\mbox{for \fhia,}\\
K^{\psi\psi^*}|W_{,\psi}|^2~~&\mbox{for \fhib,} \ecs\eeq
where $K^{\phcb\phcb^*}=K^{\Phi\Phi^*}=K^{\psi\psi^*}=1$ according
to \Eref{ktot}. Since $\vevi{S}\ll\mP$ we can also set
$K^{SS^*}\simeq1$. Taking into account the form of $W$ in
\Eref{Whi}, we find
\beq V_{\rm F0}= \bcs 4^{1 - 2 p}p^2
\sg^2\phi^{2(2p-1)}/\mst^{4(p-1)}+ \lf M^2 -
4^{-p}\phi^{2p}/\mst^{2(p-1)}\rg^2 &\mbox{for \fhia,}\\
\dq^2 \mst^{2(2-q)}q^2 \sg^2 \psi^{2(q-1)}/2 + \lf M^2 - \dq
\mst^{2 - q} \psi^q\rg^2 &\mbox{for \fhib.}\ecs \label{vfo} \eeq
whose minimization yields the SUSY vacua
\beq \vev{\sg}=0~~\mbox{and}~~ \bcs\vev{\phi}=2\lf M
\mst^{p-1}\rg^{1/p}&\mbox{for \fhia,}\\ \vev{\psi}=\lf M^2
\mst^{q-2}/\dq\rg^{1/q}&\mbox{for \fhib,} \ecs\label{vev}\eeq
with $\dq = |\Tr [T_1^q]|$ serving as a normalization factor that
depends on the representation of $\Psi$. At these vacua, the
masses of $\sg$ and $\phi$ or $\psi$ are found to be equal with
common value
\beq \label{msn} \msn= M^2\cdot\bcs \sqrt{2}p\vev{\phc}^{-1} &\mbox{for \fhia,}\\
q\vev{\psc}^{-1} &\mbox{for \fhib.}\ecs\eeq
Upon normalizing the fields w.r.t $\vev{\phi}$ and $\vev{\psi}$,
\Eref{vf} can be reformulated as follows:
\beq V_{\rm F0}=M^4\bcs \lf (1-({\phi}/{\vev{\phi}})^{2p})^2+
4p^2(\sg/\vev{\phi})^2(\phi/\vev{\phi})^{2(2p-1)}\rg&\mbox{for \fhia,} \\
\lf (1-({\psi}/{\vev{\psi}})^{q})^2+
q^2(\sg/\vev{\psi})^2(\psi/\vev{\psi})^{2(q-1)}\rg&\mbox{for
\fhib.}\ecs \label{vfnew} \eeq
Inspection of $\vfo$ shows that it features at least one
inflationary valley (with nonzero $\phi$ or $\psi$) smoothly
connected to the SUSY vacuum. Already at tree level, these valleys
display a built-in slope that drives $\sg$ towards the vacuum.
Consequently, unlike standard \cite{fhi} or shifted \cite{sh, nsh,
ssh} FHI, radiative corrections are not required to provide the
necessary slope, though they may still contribute subdominantly.
Let us, finally, note that no problematic domain walls arise from
the spontaneous breaking of the global $\mathbb{Z}_p$ or
$\mathbb{Z}_q$ symmetry in \Eref{vev}, since this breaking occurs
already during \fhi.

From here, we proceed with the analysis of the two cases, \fhia\
and \fhib, separately.

\subsubsection*{(a) Type I \fhi}

Taking the derivative of $V_{\rm F0}$ w.r.t $\phi$, we obtain
\beq  V_{\rm F0,\phi}=  4^{1 - 2 p} \mst^{2(1 -2 p)} p \phi^{2
p-3}\lf\mst^2 \phi^{2 p} \lf 2 p (2 p-1) \sg^2 + \phi^2\rg-4^p M^2
\mst^{2 p} \phi^2\rg. \label{dvf}\eeq
Assuming $\phi \ll \sg$ during \fhia, the $\phi^2$ term in the
first parenthesis can be neglected. This leads to
\beq \vevi{V_{\rm F0,\phi}}=
0~~\Rightarrow~~\vevi{\phi}=\cpa\lf\sg/\vev{\phi}\rg^{1/(1-p)}\vev{\phi}
~~\mbox{with}~~\cpa= (2p (2 p-1))^{(1/2(1 - p))},  \label{vevi}
\eeq
and $\vevi{V_{\rm F0,\phi\phi}}>0$. Substituting \Eref{vevi} into
\Eref{vf}, we obtain
\bea  \nonumber\vevi{V_{\rm F0}}&=& M^4 \lf 1 - (p-1)
\cpb(\sg/\vev{\phi})^{-2 p/(p-1)} + \cpd(\sg/\vev{\phi})^{-4 p/(p-1)} \rg , \\
&\simeq& M^4\lf 1 - (p-1) \cpb(\sg/\vev{\phi})^{-2
p/(p-1)}\rg,\label{vsm} \eea
where the $p$-dependent coefficients are defined as
\beq \cpb = \lf 2p^p (2p - 1)^{2p - 1}\rg^{1/(1 -
p)}~~\mbox{with}~~ \cpd=\cpa^{4p}. \label{cp} \eeq

\subsubsection*{(b) Type II \fhi}

Following analogous steps, we find
\beq  \vevi{V_{\rm F0,\psi}}=
0~~\Rightarrow~~\vevi{\psi}=\cqa\lf\sg/\vev{\psi}\rg^{2/(2-q)}\vev{\psi}
~~\mbox{with}~~\cqa = (q (q-1))^{1/(2 - q)}. \label{vevib} \eeq
Substituting \Eref{vevib} into \Eref{vf} yields
\bea  \nonumber\vevi{V_{\rm F0}}&=& M^4 \lf 1 -  (q - 2)
\cqb(\sg/\vev{\psi})^{-2 q/(q-2)} + \cqd(\sg/\vev{\psi})^{-4 q/(q-2)}\rg \\
&\simeq& M^4\lf 1 - (q - 2) \cqb(\sg/\vev{\psi})^{-2
 q/(q-2)}\rg,  \label{vsmb} \eea
with the $q$-dependent coefficients defined as follows
\beq \cqb = \lf q^{q}(q - 1)^{2(q -
1)}\rg^{1/(2-q)}~~\mbox{with}~~ \cqd=\cqa^{2q}. \label{cq} \eeq

\subsection{Final Form} \label{fn}

The general form of the potential that can drive the different
versions of \fhi\ can be written as
\beq\label{vhi}\Vhi=M^4\,\left(\vevi{\vfo}/M^4-{\ck}\frac{\sigma^2}{2\mP^2}+{\ckf}\frac{\sigma^4}{4\mP^4}
-{\ckx}\frac{\sigma^6}{8\mP^6}+{\ckh}\frac{\sigma^8}{16\mP^8}\right),\eeq
where $\vevi{\vfo}$ is given by \eqs{vsm}{vsmb} for \fhia\ and
\fhib\ respectively, whereas the coefficients $c_{iK}$ with
$i=2,4,6$ and $8$ depend on the SUGRA set-up in \eqs{ksh}{kn}.
Expanding the expressions of $\vf$ in \eqs{kvfsh}{vfn} as series
in powers of $\sg/\mP$ we can determine $c_{iK}$ as follows:

\begin{itemize}

\item For \shsgr\  we find
\beqs\beq  \ck=3+\bt~~\mbox{and}~~\ckf=\ckx=\ckh=0.
\label{shcon}\eeq
Given that $\bt$ is negative integer, this scenario is activated
only if we set
\beq  \ck=0~~\Rightarrow~~\beta = -3 \label{shbt}\eeq\eeqs
which totally eliminates, the quadratic SUGRA contribution to
$\eta$ parameter. On the other hand, thanks to the shift symmetry
in \Eref{ksh} all remaining SUGRA corrections in Eq.~\eqref{vhi}
vanish and, therefore, the SUSY results on the inflationary
observables reveal. The selected $\bt$ value renders the path
$\vevi{\sgb}=0$ well stabilized during \fhi\ since from
\Eref{m2sgb} with $\al=0$ (for simplicity) we obtain
\beq m^2_{\sgb}=6\lf 3 + (\sg/\mP)^2\rg
\Hhi^2\gg\Hhi^2~~\mbox{with}~~ \Hhi\simeq M^2/\sqrt{3}\mP,
\label{m2sgbH} \eeq
being the Hubble parameter during \fhi. Moreover, $\sgb$ is
sufficiently heavy and so it does not interfere with \fhi,
ensuring the consistency of our one-field inflationary scenario --
see below.

\item For \nsgr\ we find
\beqs \bea\label{c2k} c_{2K}&=&{(\al-\bt)^2\over\bt}-{2\over N},\\
\label{c4k} c_{4K}&=&{(\al-\bt)^3\over\bt^2}+{7\over2N} + {1\over N^2} + {1\over2},\\
\label{c6k} c_{6K}&=&\lf {\al^4\over\bt^3} - {3 \al^3\over\bt^2}
+{7 \al^2\over2 \bt}- 2 \al +{\bt\over2} -{2\over3} \rg \nonumber
\\ && +{1\over N}\lf
{\al^3\over\bt^2} - {5 \al^2\over2 \bt}+2 \al - {\bt\over2}-3 \rg-{1\over3N^2}, \\
\nonumber c_{8K}&=& {1\over N} \lf {\al^5\over\bt^4} - {3
\al^4\over\bt^3} + {7 \al^3\over2 \bt^2} - { 13 \al^2\over6 \bt}+
{5 \al\over6} - {\bt\over6} +{3\over8} \rg \\ && + {1\over N^2}\lf
{2 \al^4\over\bt^3} - {11\al^3\over2 \bt^2} + {11 \al^2\over2
\bt}- {5 \al\over2}  + {\bt\over2}+{13\over12}\rg \nonumber \\
&& +{1\over N^3} \lf{\al^3\over\bt^2} - {7 \al^2\over3 \bt}+ {5
\al\over3}  - {\bt\over3}-{11\over8}\rg  -{1\over12
N^4}.\label{c8k} \eea \eeqs
From Eq.~\eqref{c2k}, we see that $\ck$ can be set to zero by
constraining $N$ as a function of $\alpha$ and $\beta$, namely
\beq
\ck=0~~\Rightarrow~~N=\nc~~\mbox{with}~~\nc={2\bt\over(\al-\bt)^2}.
\label{nc}\eeq
Therefore, we are left with only two free parameters $\alpha$ and
$\beta$.
\end{itemize}

Both conditions in \eqs{shbt}{nc} consist an elegant resolution to
the infamous $\eta$ problem of FHI, since for $\alpha$ and $\bt$
of order unity -- as we show in \Sref{res} -- we can totally avoid
the relevant term.

\section{Constraining the Inflationary Dynamics} \label{cons}

The parameters of our models can be constrained by both
observational data and theoretical consistency conditions, as
described in Secs.~\ref{obs1} and \ref{obs2}.

\subsection{Observational Constraints}\label{obs1}

Our models of \fhi\ is viable only if it satisfies a number of
observational requirements. Namely:

\subparagraph{\sf\ftn (a)} The number of e-foldings underwent by
the pivot scale $\ks = 0.05/{\rm Mpc}$ during \fhi, must be
sufficient to address the standard problems of the Big Bang
cosmology \cite{hinova, plin}. Specifically,
\begin{equation}  \label{nhi}
\Ns=\:-\frac{1}{m^2_{\rm P}}\; \int_{\sgx}^{\sgf}\, d\sigma\:
\frac{V_{\rm I}}{V_{\rm I,\sg}}\simeq61.3+\frac{1-3w_{\rm
rh}}{12(1+w_{\rm rh})}\ln\frac{\pi^2g_{\rm
rh*}\Trh^4}{30\Vhi(\sgf)}+ \frac14\ln{\Vhi(\sgx)^2\over g_{\rm
rh*}^{1/3}\Vhi(\sgf)},
\end{equation}
where $\sgx$ is the field value when $\ks$ exits the horizon, and
$\sigma_{\rm f}$ marks the end of \fhi. Unlike in standard
\cite{fhi} or (semi)shifted FHI \cite{sh,nsh,ssh,su4sh}, \fhi\
ends smoothly at $\sigma = \sigma_{\rm f}$, when the slow-roll
conditions break down. Concretely, $\sigma_{\rm f}$ is determined
by
\beq \label{slow} {\sf\ftn max}\{\epsilon(\sigma_{\rm
f}),|\eta(\sgf)|\}=1,~~\mbox{where}~~ \epsilon\simeq{m^2_{\rm
P}\over2}\left(\frac{V_{\rm I,\sg}}{V_{\rm
I}}\right)^2~~\mbox{and}~~\eta\simeq m^2_{\rm P}~\frac{V_{\rm
I,\sg\sg}}{V_{\rm I}}\cdot \eeq
In the last part of \Eref{nhi} we assume that \fhi\ is followed in
turn by an oscillatory phase (due to $S$) with mean
equation-of-state parameter $w_{\rm rh}$, radiation and matter
domination. Also $\Trh$ is the reheat temperature after \fhi.
Given that $w$ is found by the standard formula \cite{lect}
\beq w_{\varphi}=(n-2)/(n+2)~~\mbox{for a power-law potential}~~
\varphi^n, \label{wrh}\eeq
we take for our numerics $w_{\rm rh}=w_{S}=0$ which corresponds
precisely to $n=2$, since $\vfo$ in \Eref{vfo} displays a
quadratic dependence on $S$ for $\phi$ close to its v.e.v in
\Eref{vev}. On the other hand, the oscillations of the Higgs
fields give rise to a stiff component of the total matter
energy-density -- with $w_{\rm H}>1/3$ since $n>4$ -- which is
overshadowed by that of $S$ \cite{bernal}. This intuitive picture
is verified and further explained through a numerical example
presented in Appendix~\ref{lepto}. Motivated by implementations
\cite{sm1, smii, zubair} of non-thermal leptogenesis \cite{lept},
which may follow \fhi, we set $\Trh\simeq(10^8 -10^9)~\GeV$.
Although not crucial for the resulting magnitude of $\Ns$, we
mention that we take for the energy-density effective number of
degrees of freedom $g_{\rm rh*}=228.75$ inspired by the MSSM
spectrum. With these choices, we obtain $ \Nhi\simeq 48.2-49.5$.


\subparagraph{\sf\ftn (b)} The amplitude $A_{\rm s}$ of the
curvature perturbation power spectrum generated by $\sigma$ during
\fhi\ must be consistent with CMB observations \cite{actin}, i.e.
\begin{equation} \label{prob}
A_{\rm s}= \frac{1}{12\, \pi^2 m^6_{\rm P}}\; \left.\frac{V_{\rm
I}^{3}(\sgx)}{|V'_{\rm I}(\sgx)|^2}\right.\simeq2.1326\cdot
10^{-9}.
\end{equation}

\subparagraph{\sf\ftn (c)} The remaining key observables are the
scalar spectral index $n_{\rm s}$, its running \as, and the
tensor-to-scalar ratio $r$. These are obtained from the standard
expressions
\beq \label{ns}  \ns = 1-6\epsilon_\star\ + \ 2\eta_\star, \,
\as={2}\left(4\eta_\star^2-(\ns-1)^2\right)/3-2\xi_\star~~\mbox{and}~~
r=16\epsilon_\star,\eeq
where $\xi \simeq m_{\rm P}^4 (V'_{\rm I} V'''_{\rm I})/V_{\rm
I}^2$, and all quantities with a subscript $\ast$ are evaluated at
$\sigma = \sigma_\ast$. As regards $\ns$ we consider implications
to our models taking into account  separately \eqs{nsact}{nsspt}.
The remaining observables must lie within the following
approximate ranges \cite{actin}:
\beq \label{data} \as  = 0.0062\pm0.0104~~\Rightarrow~~
-0.004\lesssim \as \lesssim 0.017 ~~\mbox{and}~~r\lesssim0.038,
\eeq at 95\% c.l.
Note that both observables above are compatible with their values
in our context, as we show in \Sref{res}.

\subsection{Theoretical Considerations} \label{obs2}

From the theoretical perspective, the viability of our model can
be further refined by imposing the following requirements:

\subparagraph{\sf\ftn (a)}  \textit{Unification Constraint}. If
the gauge group $\Ggut$ contains non-Abelian factors beyond those
of MSSM, the mass $M_A$ of the lightest gauge boson $A_\mu$ in the
SUSY vacuum -- see \Eref{vev} -- must be consistent with gauge
coupling unification within MSSM, namely
\beq \label{Mgut} M_{A}\simeq2 \cdot 10^{16}~\GeV\>\mbox{where}\>
M_{A} = \bcs {g \vev{\phc}} & \mbox{for \fhia,} \\ {g  n_{\rm
G}\vev{\Psi}} & \mbox{for \fhib,}\ecs\eeq
where $g \simeq 0.7$ is the unified gauge coupling constant, and
$n_{\rm G}$ is a group-theoretic normalization factor depending on
$\Ggut$. For example, in the case $\Ggut = SU(5)$, one finds
$n_{\rm G} = \sqrt{5/6}$.

\subparagraph{\sf\ftn (b)} \textit{Boundedness of $V_{\rm I}$.}
The inflationary potential must be bounded from below to avoid the
possibility of a disastrous runaway of the system to infinite
values of the inflaton field. This requirement also facilitates
the possibility that the system may eventually undergo an
inflationary expansion under generic initial condition.

\subparagraph{\sf\ftn (c)} \textit{Convergence of $V_{\rm I}$.}
The expansion of $\Vhi$ in \Eref{vhi} should converge at least up
to the relevant field values $\sigma \sim \sigma_\ast$. This is
guaranteed if, for such values, each successive term in the
aforementioned expansion is smaller than the preceding one. In
practice, this requirement can be satisfied if our results remain
immune from the contributions with coefficients $c_{iK}$ with
$i>6$.

\subparagraph{\sf\ftn (d)} \textit{Monotonicity of $V_{\rm I}$.}
Depending on the values of the coefficients $c_{iK}$ in
\Eref{vhi}, the potential $V_{\rm I}$ may either remain a
monotonic function of $\sigma$ or develop a local minimum and
maximum. In the scenarios of interest here, $V_{\rm I}$ turns out
to be monotonic, avoiding the complications of metastable extrema
-- see \Sref{res3}.

\section{Results}\label{res}

The discussed scenarios of \fhi\ depend on the following
parameters
$$M,~\mst~~\mbox{and}~~p~[q]~~\mbox{for \fhia\ [\fhib] and}~~\al,~\bt~~\mbox{only for \nsgr.}$$
Recall that the resolution of the $\eta$ problem forces us to
determine $\bt$ from \Eref{shbt} for \shsgr\ and $N$ as a function
of $\al$ and $\bt$ via \Eref{nc} for \nsgr. The unification
condition of \Eref{Mgut} by virtue of \Eref{vev} allows us to
obtain $\mst$ in terms of $M$ as follows
\beq \mst=\bcs \lf\vev{\Phi}^p/M\rg^{1/(p-1)}&\mbox{for \fhia,} \\
\lf\vev{\Psi}\dq/n_{\rm G}M^{2/p}\rg^{p/(p-2)} &\mbox{for \fhib,}
\ecs \eeq
where $\dq=2^{-q} 15^{-q/2} \lf3\cdot 2^q (-M)^q + 2\cdot3^q
M^q\rg$ -- note that for \fhib\ we confine ourselves in the case
with $\Ggut=SU(5)$. Moreover, $M$ can be determined by enforcing
\Eref{prob} -- note that \Eref{nhi} constrains $\sgx$ which is an
internal parameter of the inflationary setting. Therefore, the
number of the free parameters of our models can be reduced by two.

\subsection{\shsgr} \label{res1}

In the context of \shsgr\ $\Vhi$ is given by the general form of
\Eref{vhi} with imposition of the conditions in \eqs{shcon}{shbt}.
As a consequence, $\Vhi$ coincides with $\vevi{\vfo}$ in
\Eref{vsm} or (\ref{vsmb}) for \fhia\ or \fhib\ respectively and
so the observables are totally identical to those obtained within
rigid SUSY, i.e., they only depend on the parameters $p$ or $q$
for \fhia\ or \fhib\ respectively. Our results are summarized for
each type of \fhi\ separately below.

\subsubsection*{(a) Type I \fhi}

In this case the slow-roll parameters can be found by inserting
$\Vhi$ as described above into \Eref{slow} with result
\beq \label{epa}  \ep_0 = 2p^2\cpb^2 \lf\frac{\mP}{\sg}\rg^2
\lf\frac{\sg}{\vev{\phi}}\rg^{4 p/(1-p)}~~\mbox{and}~~ \eta_0 =
\frac{4 \cpb  (1 - 3 p) p}{(p-1)} \lf\frac{\mP}{\sg}\rg^2
\lf\frac{\sg}{\vev{\phi}}\rg^{2 p/(1 - p)}. \eeq
The number of e-foldings $\Ns$ is found by applying the left-most
expression in \Eref{nhi} as follows
\beq \Ns=\frac{(p-1) \sgx^{4 + 2/(p-1)} \vev{\phi}^{2 p/( 1-p)}}{4
\cpb \mP^2 p (2 p-1)}~~\Rightarrow~~\sgx=2^{ \frac{(p-1)}{2 p-1}}
\lf \frac{(p-1) \vev{\phi}^{-2 p/(p-1)}}{\cpb \mP^2 \Ns p (2
p-1)}\rg^{\frac{1 - p}{4 p-2}}.\label{sgxa} \eeq
Solving \Eref{prob} w.r.t $M$  we find
\beq  M\simeq2\mP^{3/2} \sqrt{\pi p\cpb \sqrt{3\As}}\sgx^{-1/2}
\lf\vev{\phi}/\sgx\rg^{p/(p-1)},\label{Ma} \eeq
where $\sgx$ is given by \Eref{sgxa} -- its explicit replacement
in the expression above yields a rather lengthly final result.
Substituting also \Eref{sgxa} into \Eref{ns} we arrive at the
predictions for $\ns$ and $\as$ which become
\beq \label{nsa}  n_{\rm s0} = 1 - {3 p-1 \over 2
p-1}\frac{1}{\Ns}~~\mbox{and}~~a_{\rm s0}=- {3 p-1 \over 2
p-1}\frac{1}{\Ns^2} \eeq
whereas $r$ remains rather suppressed. The expression of $\ns$ is
in agreement with the one found in \cref{okada}. Numerical and
more precise values for the quantities above are accumulated in
\Tref{tab} for $p = 2, 3$ and $4$. We observe that $\sgx, \sgf$
and $\mst$ are of order $\mstr$ whereas $M$ and $\msn$ are of
order $1~\YeV$ and $0.1~\YeV$ increasing slightly with $p$ --
recall that $1~\YeV=10^{15}~\GeV$. The $\ns$ values are
impressively consistent with the \spt\ data in \Eref{nsspt} and
marginally disfavored by the \actc\ data in \Eref{nsact}
independently from $p$. On the other hand, $\as$ and $r$ safely
satisfy the observational bounds in \Eref{data}. Lastly, the
identification of $\mst$ with $\mP$ is not possible since $\kp$
defined in \Eref{kp} violates the perturbative limit.

\renewcommand{\arraystretch}{1.2}
\begin{table}[!t]
\begin{center}
\begin{tabular}{|l|lll||lll|}
\hline
{\sc Cases}&\multicolumn{3}{|c||}{{\sc Type} I \fhi}&\multicolumn{3}{|c|}{{\sc Type} II \fhi}\\
\hline
&$p=2$&$p=3$&$p=4$&$q=3$&$q=5$&$q=7$ \\ \hline \hline
$\sgx/10^{17}~\GeV$ & $2.65$&$3.32$&$3.47$&$1.49$&$2.64$&$2.95$ \\
$\mst/\mstr$ & $1.57$&$0.27$&$0.16$&$8.5$&$0.14$&$0.07$ \\
\hline
$\kp$&\multicolumn{3}{|c||}{$>\sqrt{4\pi}$}&$0.57$&\multicolumn{2}{|c|}{$>\sqrt{4\pi}$} \\
\hline
$M/10^{15}~\GeV$& $1$&$1.3$&$1.35$&$0.7$&$1.1$&$1.2$ \\
$\sigma_{\rm
f}/10^{17}~\GeV$&$1.34$&$1.45$&$1.42$&$0.85$&$1.22$&$1.24$ \\
$N_{\rm I*}$ & $48.8$&$49.3$&$49$&$48.6$&$48.9$&$48.9$ \\ \hline
$n_{\rm s}$ & $0.966$&$0.968$&$0.968$&$0.964$&$0.967$&$0.968$ \\
$-\alpha_{\rm s}/10^{-4}$ & $6.7$&$6.4$&$6.3$&$7.4$&$6.7$&$6.4$ \\
$r/10^{-6}$ & $1.1$&$2.4$&$3.8$&$0.197$&$1.3$&$2.1$ \\ \hline
$\msn/10^{14}~\GeV$ & $1.1$&$2.4$&$3.6$&$0.45$&$1.94$&$3.4$\\
\hline
\end{tabular}
\end{center}
\caption[]{\sl\small Input and output parameters compatible with
Eqs.~(\ref{nhi}), (\ref{prob}) and (\ref{Mgut}), within shSUGRA
(or SUSY) for \fhia\ or \fhib\ and selected $p$ or $q$ values
respectively.}\label{tab}
\end{table}

\renewcommand{\arraystretch}{1.}

\subsubsection*{(b) Type II \fhi}

For \fhib, a similar analysis yields for the slow-roll parameters
in \Eref{slow}
\beq \label{epb}  \ep_0 = 2q^2\cqb^2 \lf\frac{\mP}{\sg}\rg^2
\lf\frac{\sg}{\vev{\psi}}\rg^{4 q/(2 -q)}~~\mbox{and}~~ \eta_0 =
\frac{2 \cqb (2 - 3 q) q }{(q-2)} \lf\frac{\mP}{\sg}\rg^2
\lf\frac{\sg}{\vev{\psi}}\rg^{2 q/(2-q)}, \eeq
while $\sgx$ as a function of $\Ns$ can be found as follows
\beq \label{sgxb}  \Ns\simeq \frac{(q-2) \sgx^{4 + 4/(q-2)}
\vev{\psi}^{2q/(2-q)}}{8 \cqb \mP^2
(q-1)q}~~\Rightarrow~~\sgx=2^{3 (q-2)\over 4 (q-1)} \lf\frac{(q-2)
\vev{\psi}^{-2 q/(q-2)}}{\cqb \mP^2 \Ns (q-1) q}\rg^{\frac14
{(2-q)\over(q-1)}}. \eeq
Also the mass scale $M$ reads
\beq M = 2 \mP^{3/2} \sqrt{\pi q \cqb \sqrt{3\As} } \,
\sgx^{-1/2}\lf\vev{\psi}/\sgx\rg^{q/(q-2)}, \label{Mb} \eeq
where $\sgx$ is given by \Eref{sgxb}. The corresponding spectral
observables are
\beq \label{nsb}  n_{\rm s0} = 1 - {3 q-2 \over 2 (q-
1)}{1\over\Ns} ~~\mbox{and}~~ a_{\rm s0}  =- {3 q-2
\over2(q-1)}{1\over\Ns^2}. \eeq
The results for $\ns$ and $\as$ above can be derived by the
corresponding ones in \Eref{nsa} replacing there $p=q/2$.
Numerical and more precise values for the quantities above are
accumulated in \Tref{tab} for $q = 3, 5$ and $7$ with similar
observational results as those in \fhia. Since even values of
$q>2$ yield almost identical results with \fhia\ with $p=q/2$, we
restrict our discussion to odd values of $q$. In particular, the
simplified version of \fhib\ with $\kappa < \sqrt{4\pi}$ is viable
only for $q = 3$, where $\mst$ can be identified with the Planck
mass $m_P$. In other cases, this identification requires
$\kappa > \sqrt{4\pi}$.

\subsection{\nsgr}\label{res2}

In the context of \nsgr, $\Vhi$ is given by the general form of
\Eref{vhi} with the coefficients in \Eref{c2k} -- (\ref{c8k})
after the imposition of the conditions in \Eref{nc}. In practice,
only $c_{4K}$ from the coefficients above plays a crucial role for the
determination of the inflationary observables. For both \fhia\ or
\fhib\ the slow-roll parameters read
\beq \ep\simeq\lf\ep_0^{1/2} +{c_{4K}\sg^3}/{\sqrt{2}\mP^3}
\rg^2~~\mbox{and}~~\eta\simeq\eta_0+3c_{4K}(\sg/\mP)^2,\eeq
where $\ep_0$ and $\eta_0$ are given by \Eref{epa} or (\ref{epb})
for \fhia\ or \fhib\ respectively. Inserting $\ep$ into the
leftmost part of \Eref{nhi} we can compute $\Ns$ with result
\beq \Ns\simeq\frac{\mP^2}{2c_{4K}\sgx^2} \bcs {}_2F_1\lf
1;\frac{1-p}{3p-2};\frac{2p-1}{3p-2};-\frac{c_{4K}\sgx^{2(3p-4)/(p-1)}}
{2p\cpb\mP^4\vev{\phi}^{2p/(p-1)}}\rg & \mbox{for \fhia,}\\[4mm]
{}_2F_1\lf
1;\frac{q-2}{3q-4};\frac{2(q-1)}{3q-4};-\frac{c_{4K}\sgx^{2(3q-4)/(q-2)}}
{2q\cqb\mP^4\vev{\psi}^{2q/(q-2)}}\rg& \mbox{for \fhib.} \ecs\eeq
Here ${}_2F_1$ is the Gauss hypergeometric function. Due to these
complicate forms of $\Ns$ we are not able to solve the expressions
above w.r.t $\sgx$ and pursue our analytical approach as in
\Sref{res1}. Assuming, however, that the $\sgx$ values do not
deviate a lot from their values in \Eref{sgxa} or (\ref{sgxb}) for
\fhia\ or \fhib\ respectively we can provide analytical
approximations that capture the essential behavior of $\ns$.
Namely we find
\begin{equation}
\label{nsn} \ns \simeq n_{\rm s0} +3c_{4K}\bcs
4\lf\vev{\phi}/{\mP}\rg^{2p/(2p-1)}
\lf\Ns\cpb p(2p-1)/(p-1)\rg^{(p-1)/(2p-1)}& \mbox{for \fhia,}\\[4mm]
2^\frac{5q-8}{2(q-1)}\lf\vev{\psi}/{\mP}\rg^{q/(q-1)}\lf\Ns\cqb
q(q-1)/(q-2)\rg^{(q-2)/2(q-1)}& \mbox{for \fhib.} \ecs
\end{equation}
Clearly, we can obtain an elevation of the resulting $\ns$ w.r.t
$n_{\rm s0}$ provided that $c_{4K}$ in \Eref{c4k} is positive.
Indeed, taking into account \Eref{nc} it is not difficult to show
that $c_{4K}>0$ for $|N|<7$.

\begin{figure}[!t]
\centering
\begin{minipage}{0.48\textwidth}
   \includegraphics[width=\linewidth]{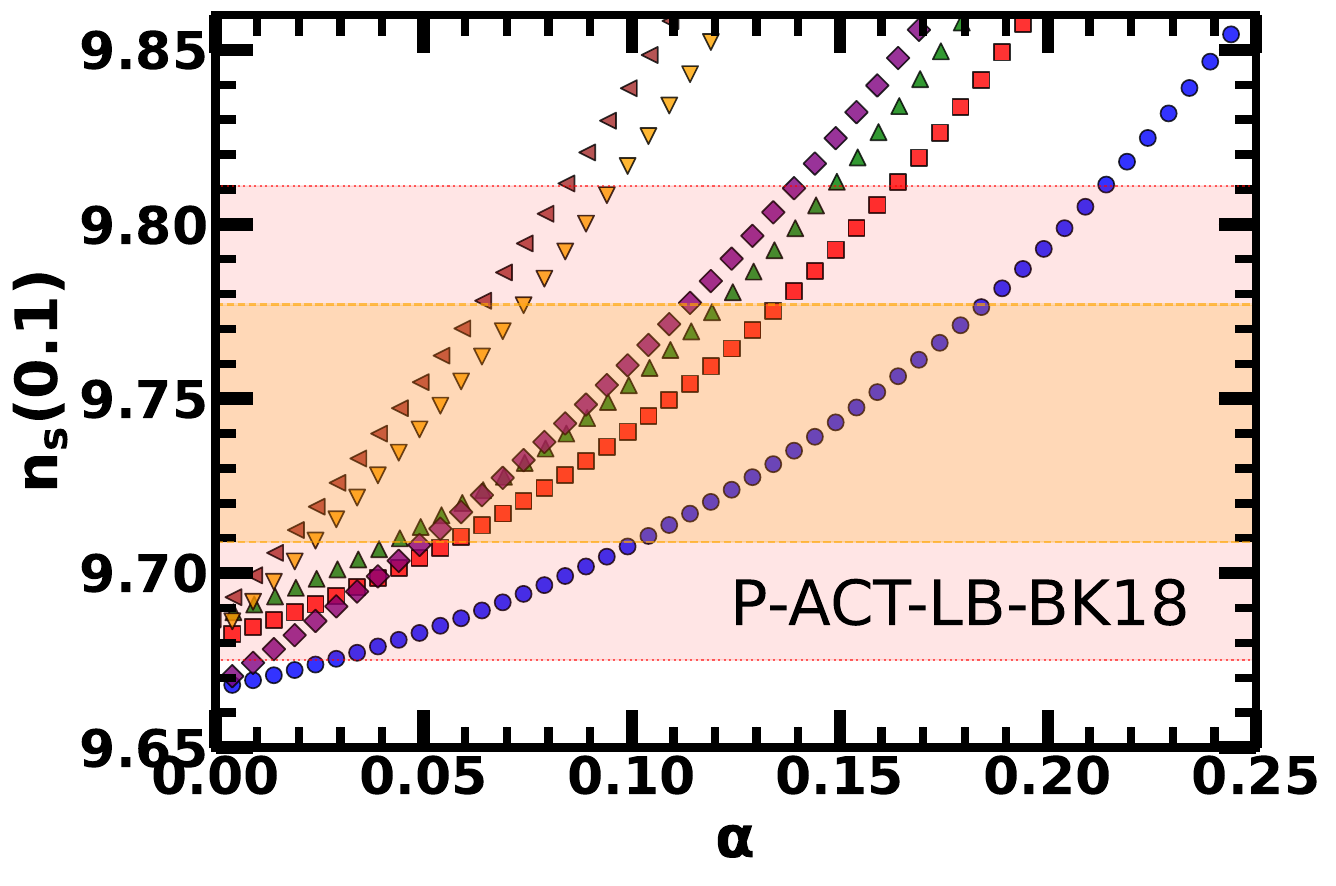}
\end{minipage}\hfill
\begin{minipage}{0.48\textwidth}
   \includegraphics[width=\linewidth]{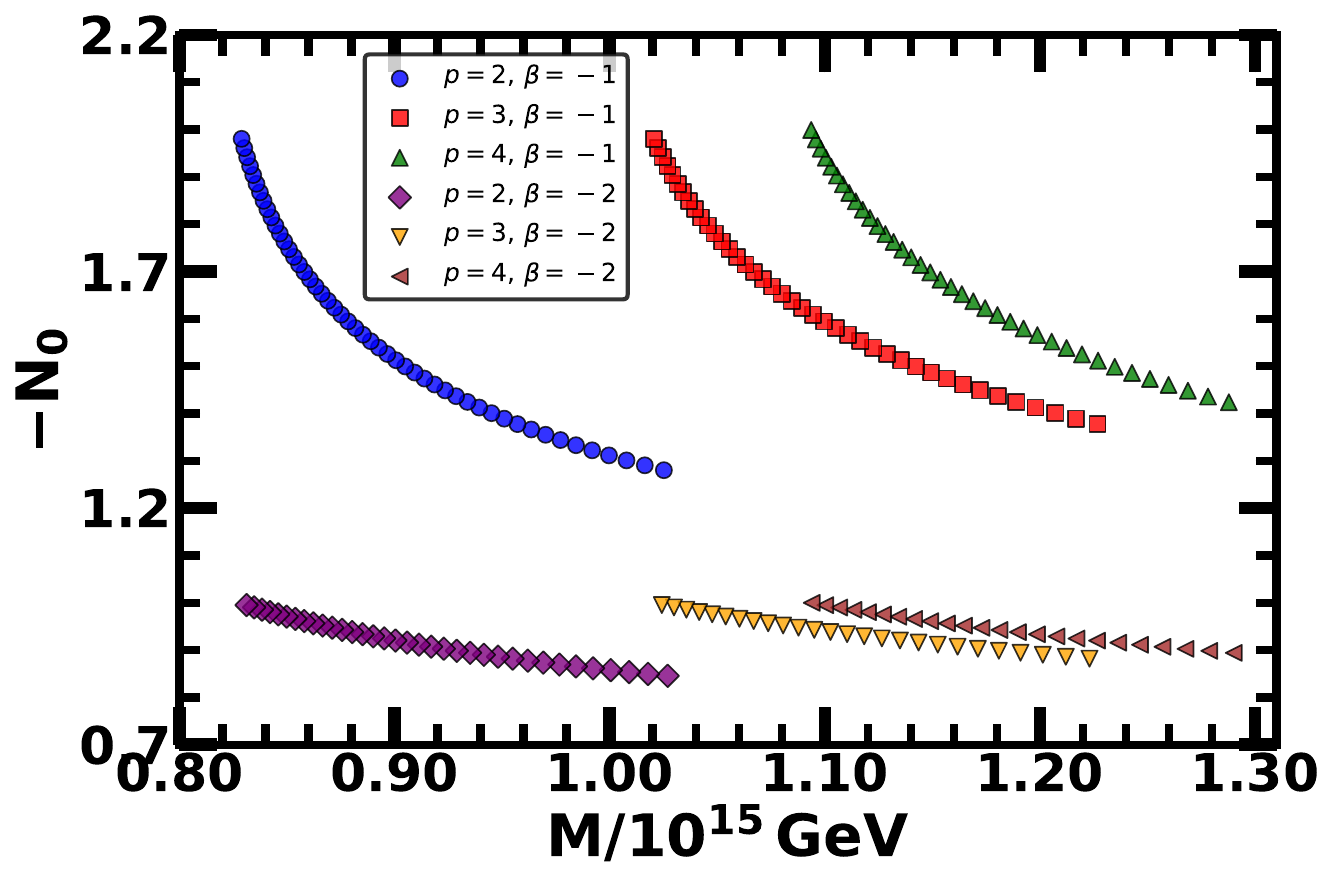}
\end{minipage}
\hfill \caption[]{\sl\small  Values of $\ns$ allowed by
Eqs.~(\ref{nhi}), (\ref{prob}) and (\ref{Mgut}) versus $\al$ for
\fhia, $\bt=-1$ or $-2$ and $p=2,3$ and $4$ -- the marginalized
joint $68\%$ [$95\%$] c.l. regions from \actc\ data are depicted
by the dark [light] shaded contours (left plot). Resulting values
of $(-\nc)$ versus $M$ for the same $\bt$ and $p$ (right plot).
Shown is also the applied color coding in both plots for the
various $\bt$ and $p$ values in the legend of the right
plot.}\label{fig1}
\end{figure}

\begin{figure}[ht]
\centering
\begin{minipage}{0.48\textwidth}
    \includegraphics[width=\linewidth]{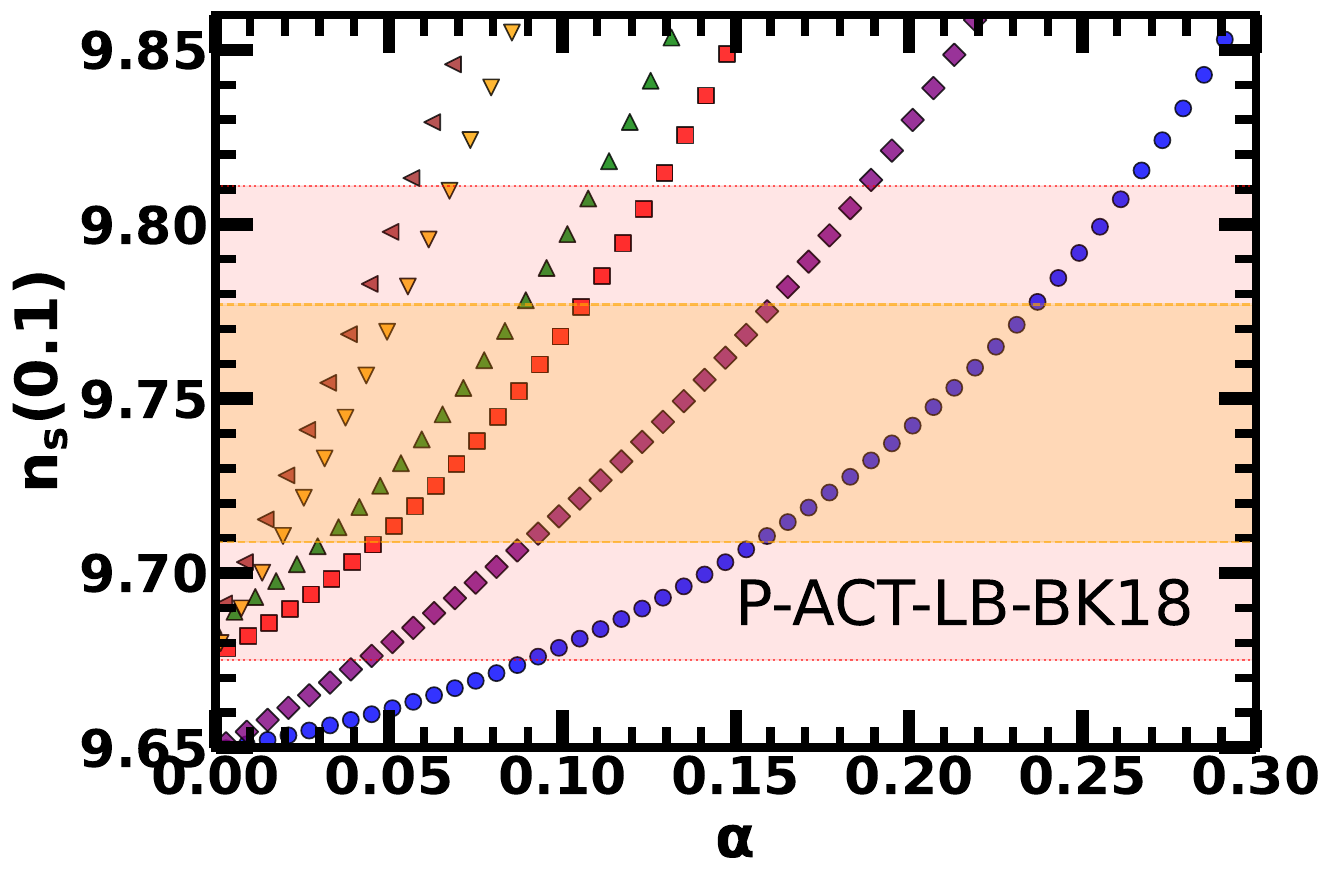}
\end{minipage}\hfill
\begin{minipage}{0.48\textwidth}
    \includegraphics[width=\linewidth]{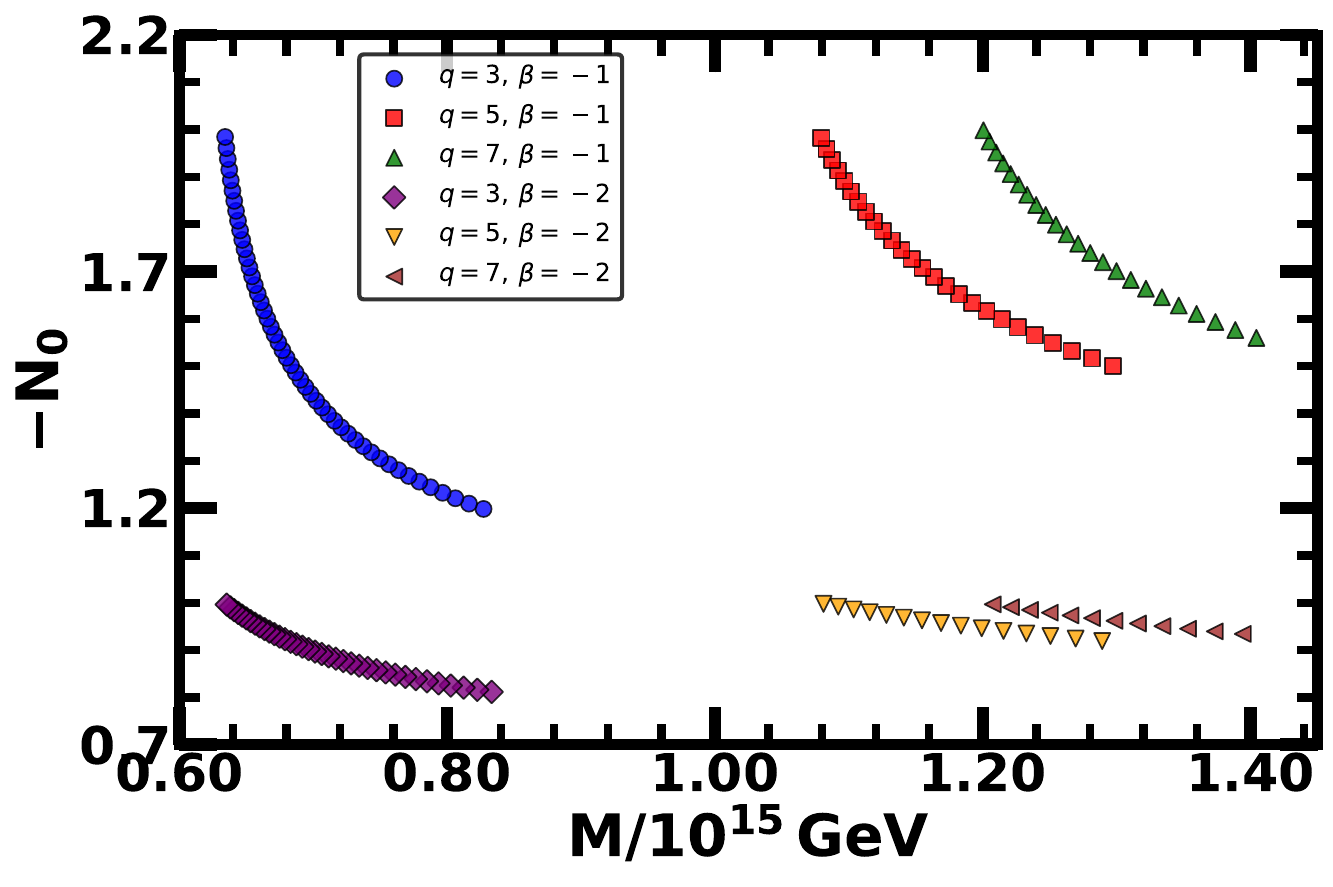}
\end{minipage}
\hfill\caption[]{\sl\small The same as in \Fref{fig1} but for
\fhib\ and $q=3,5$ and $7$ (instead of $p$).  } \label{fig2}
\end{figure}

\begin{figure}[ht]
\centering
\begin{minipage}{0.48\textwidth}
\includegraphics[width=\linewidth]{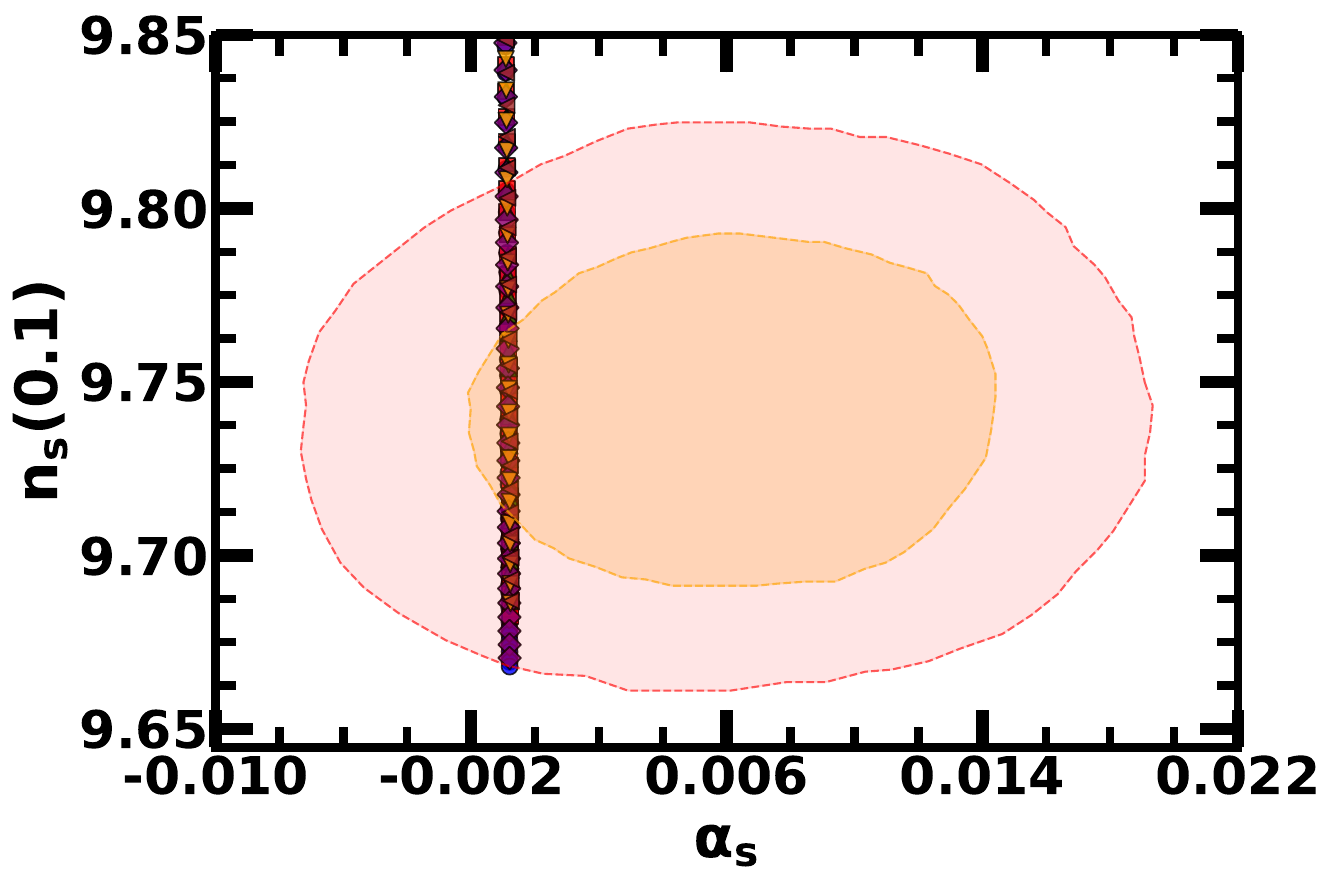}
\end{minipage}
\begin{minipage}{0.48\textwidth}
\includegraphics[width=\linewidth]{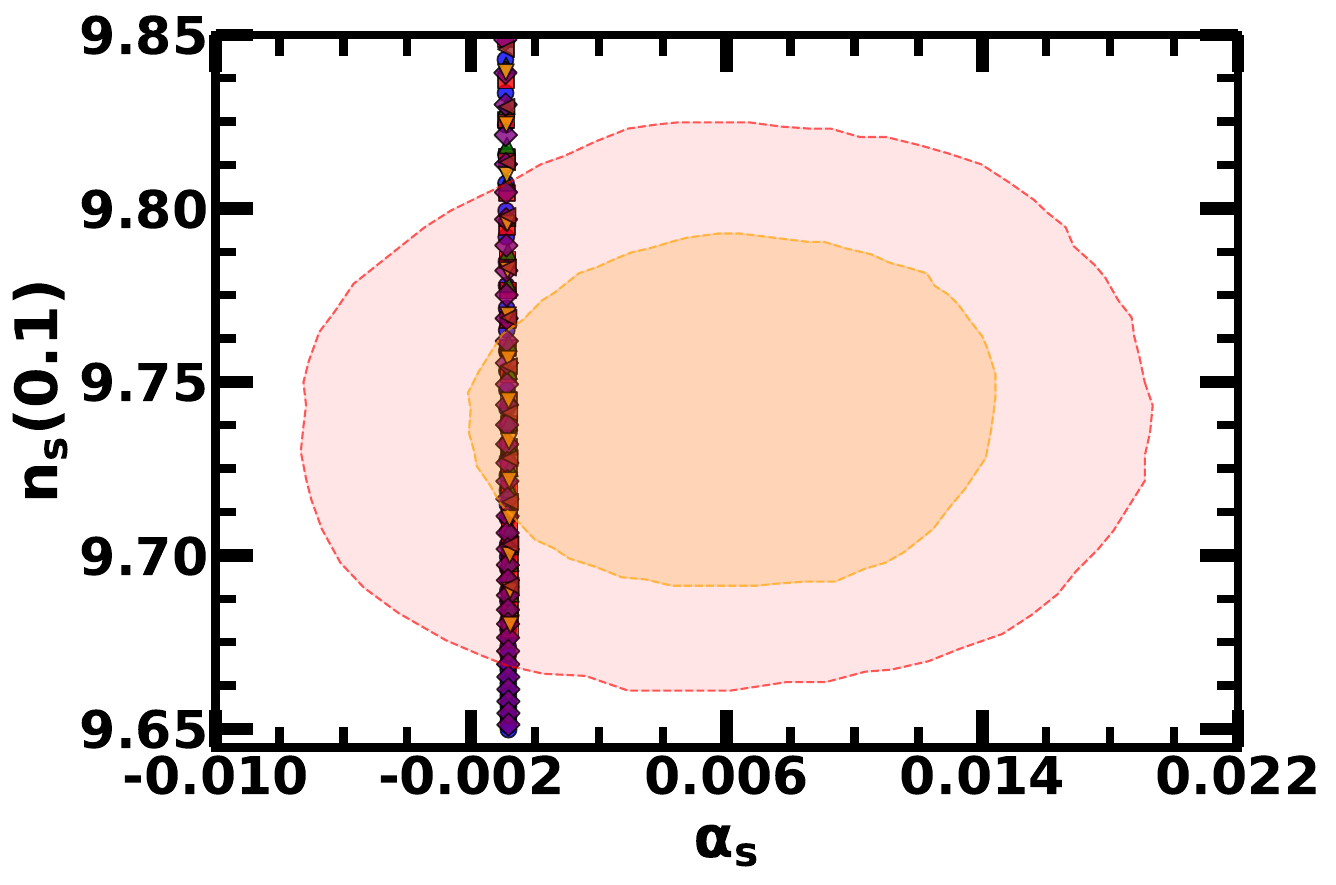}
\end{minipage}
\hfill \caption[]{\sl\small Values of $\ns$, allowed by
Eqs.~(\ref{nhi}), (\ref{prob}) and (\ref{Mgut}), versus $\as$ for
\nsgr and \fhia\ [\fhib] (left [right] plot) corresponding to the
parameter choices discussed in Fig.~\ref{fig1} [Fig.~\ref{fig2}].
The marginalized joint 68\% [95\%] c.l. regions derived from
\actb\ data are indicated by the dark [light] shaded contours.
}\label{fig4}
\end{figure}

This insight can be explicitly verified by carrying out a precise
numerical analysis taking into account all the coefficients in
\Eref{c2k} -- (\ref{c8k}). We use as input parameters $\al$ and
$\bt=-1$ or $-2$ for selected values of $p$ (\fhia) or $q$
(\fhib). Our results are shown in \Fref{fig1} for \fhia\ and
\Fref{fig2} for \fhib. In the left panel of each figure we present
the allowed values of $\ns$ versus $\al$ whereas in the right
panels we display the corresponding values of $(-\nc)$ as a
function of the mass scale $M$. The inner and outer shaded regions
represent the marginalized joint 68\% and 95\% c.l. contours from
the \actc\ data. The choices of $p$ and $q$ are given in the right
plots. For both types of \fhi\ we observe that there is a plethora
of $\al$ values which lead to \actc-compatible $\ns$ for both
chosen $\bt$. Their magnitudes are less than $0.3$ and decrease as
$|\bt|$ increases. Also $|\nc|$ decreases below unity as $|\bt|$
increases. Therefore, low $p, q$ and $|\bt|$ values are better
motivated from the point of view of naturality. The relevant mass
scale $M$ turns out to be of order $1~\YeV$. E.g., fixing $\bt=-1$
we obtain
\beqs\bea \label{resa}2/125<\alpha<477/2000~~\mbox{with}~~0.79 <M /\YeV <0.93 & \mbox{for \fhia\ and $p=2$} ; \\
41/500<\al<67/250~~\mbox{with}~~0.65 <M /\YeV <0.79&\mbox{for
\fhib\ and $q=3$}. \label{resb} \eea \eeqs
In \Fref{fig4} we also display the $\as$ predictions of \fhia\
(left plot) and \fhib\ (right plot) as a function of $\ns$ for the
same $\bt$ and $p$ or $q$ values used in \Fref{fig1} or
\ref{fig2}. The consideration of the $\ns-\as$ correlations
reveals that the negligibly small $|\as|$ values obtained in our
setup are consistent not only with its $95\%$ c.l. allowed
approximate margin in \Eref{data}, but also with the $68\%$ c.l.
allowed margin for $0.971\lesssim\ns\lesssim0.976$. Note that
similar $\as$ (and $r$) values are encountered within \shsgr\ too.

Let us emphasize, finally, that the \spt\ range in \Eref{nsspt}
can be obviously accommodated within \nsgr\ too. However, the
obtained $\al$ and $\nc$ values are expected to be somehow
suppressed and therefore we consider that this framework is more
appropriate to fit the \actc\ results in \Eref{nsact}. Also, we
can obtain acceptable results even for $\al=\bt=0$, i.e., without
the presence of $\Z$ and $\K$ in \Eref{kn} at the cost of a very
high (positive) $N\sim 800$. Although this value is perfectly
acceptable within SUGRA, it has no theoretical motivation and so
it can be considered as less natural.

\subsection{Monotonicity of \Vhi}\label{res3}

\begin{figure}[!t]
\begin{minipage}{0.48\textwidth}
  \includegraphics[width=\linewidth]{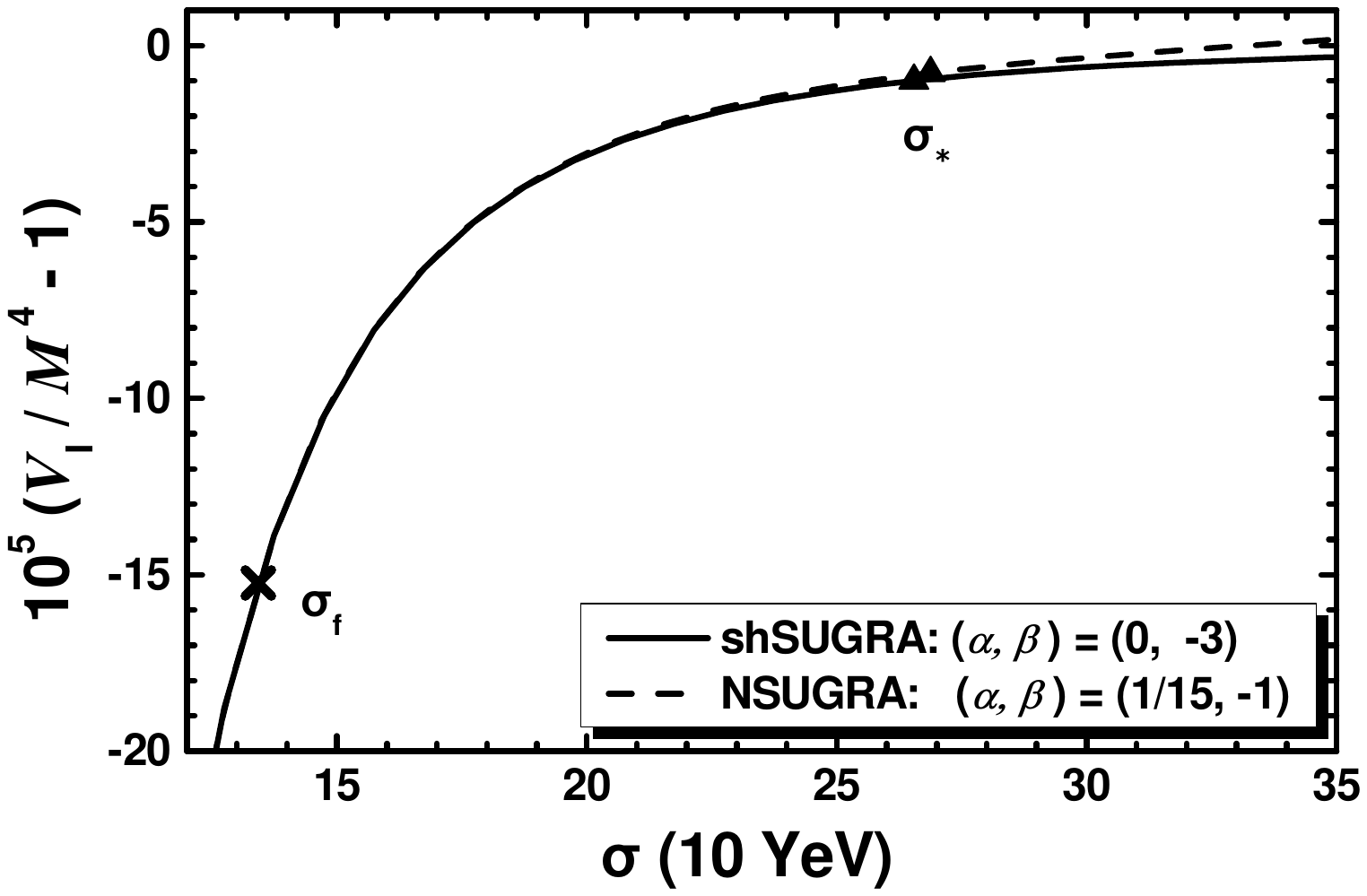}
\end{minipage}\hfill
\begin{minipage}{0.48\textwidth}
    \includegraphics[width=\linewidth]{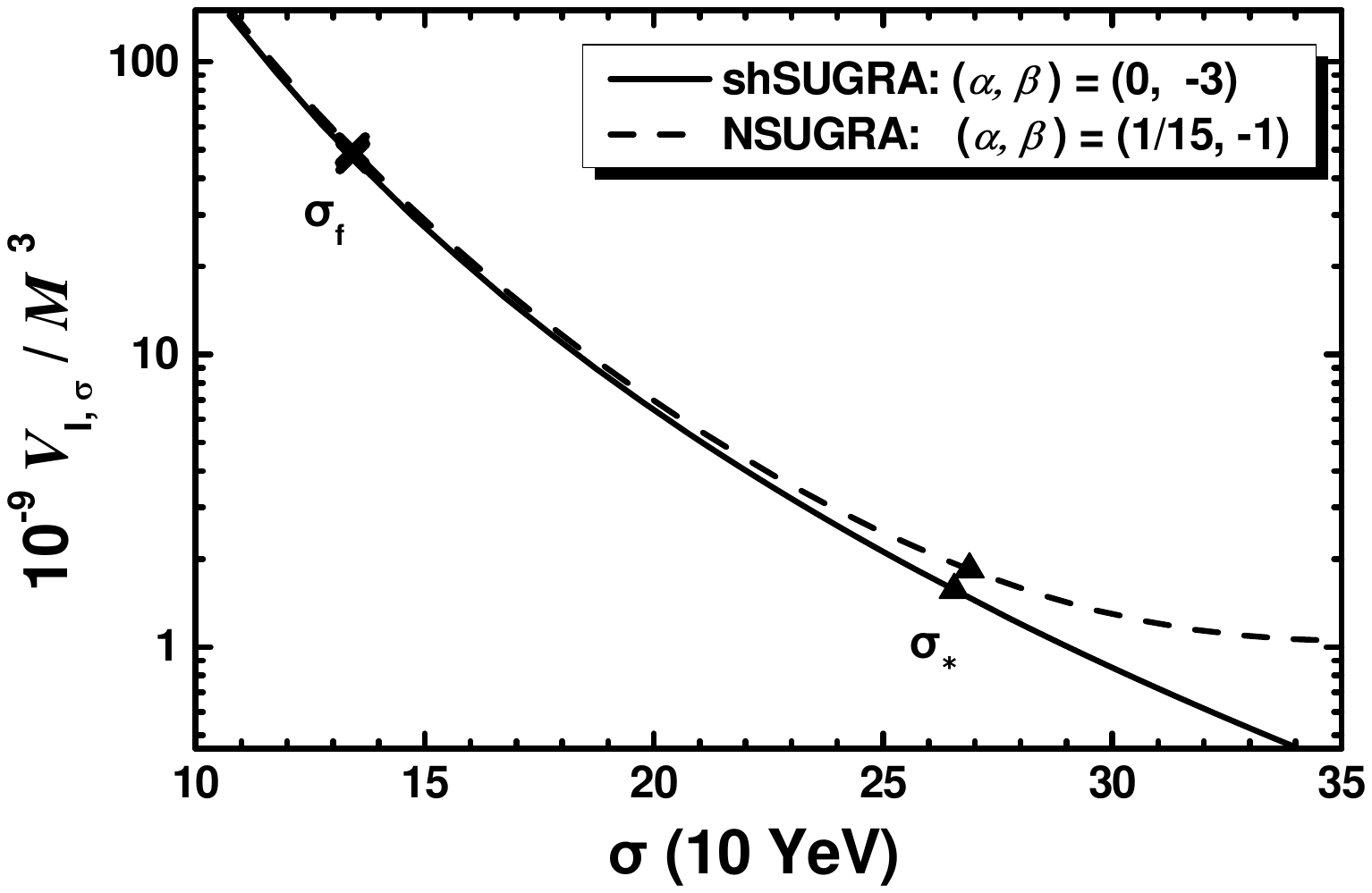}
\end{minipage}
\hfill \caption[]{\sl\small The variation of $\Vhi$ in \Eref{vhi}
(left panel) and $V_{\rm I,\sg}$ (right panel) as functions of
$\sg$ for \fhia\ with $p=2$, \shsgr\ $\al=0$ and $\bt=-3$
resulting to $\ns=0.967$ (solid line) or \nsgr\ $\al=1/15$ and
$\bt=-1$ resulting to $\ns=0.972$ (dashed line). The values of
$\sgx$ and $\sgf$ are also depicted. }\label{fig3}
\end{figure}

An outstanding feature of our proposal is that $\Vhi$ in
\Eref{vhi} remains monotonic during \fhi\ for both SUGRA settings.
To highlight it, we display the variation [the derivative w.r.t
$\sg$] of $\Vhi$ as a function of $\sg$ in the left [right] plot
of \Fref{fig3}. We focus on \fhia\ with $p=2$. For \shsgr\ (solid
lines) we use the parameters shown in the column with $p=2$ in
\Tref{tab} whereas for \nsgr\ (dashed lines) we use $\al=1/15$ and
$\bt=-1$ resulting to $\nc\simeq-1.76$ and $\ns=0.9723$. The
values of $\sgx$ and $\sigma_{\rm f}$ are also depicted. Note that
$\sgx$ and $\sgf$ coincide for the two presented lines and so our
assumption for the derivation of \Eref{nsn} is justified. Most
importantly, it is clear that $\Vhi$ remains  a monotonically
increasing function of $\sg$ in both cases since $V_{\rm
I,\sg}>0$. So, unnatural restrictions on the initial conditions
for \fhi\ due to the appearance of a maximum and a minimum of
$\Vhi$ can be avoided.

\newpage

\section{Conclusions} \label{con}

We have constructed and analyzed two types (I and II) of smFHI
(i.e., smooth F-term hybrid inflation) depending on the
representation of the Higgs involved fields -- see \Eref{Whi}. A
key ingredient of our proposal is that the parameters of the
superpotential in \Eref{Whi} are constrained to values that ensure
compatibility with the well-established requirement of gauge
coupling unification within the MSSM according to\Eref{Mgut}. This
provides a robust theoretical foundation, anchoring our
inflationary scenario in a broader particle-physics framework.

However, when confronted with the latest observational constraints
of \actc\ and \spt\ data, the model is found to be incompatible
within mSUGRA -- see Appendix~\ref{msgr}. To control SUGRA
corrections, we incorporate the impact on the inflationary
dynamics from a stabilized modulus. It appears only in the \Ka of
our set-up through the functions $\Z$ and $\K$ in \Eref{kh} with
parameters $\al$ and $\bt$ inspired by superstring and
D-brane-motivated models. We also considered two classes of \Kas\
for the inflaton sector in \eqs{ksh}{kn} which define the SUGRA
settings, \shsgr\ and \nsgr\ respectively. In \shsgr, the
combination of the shift symmetry for the inflaton-sector together
with the adjustment of $\bt$ leads to predictions that coincide
with the SUSY version of \fhi\ and remain consistent with \spt\
data. In \nsgr, the conjunction of the inflaton hyperbolic
geometry with the variation of $\al$ and $\bt$ assist to tame the
$\eta$ problem and achieve full consistency with \actc\ data --
see Figs. \ref{fig1} and \ref{fig2}. It is gratifying that our
solutions are obtained keeping the inflationary potential
monotonic -- see \Fref{fig3}.

Let us, finally, note that a complete inflationary scenario must
also account for the transition to a radiation-dominated universe
and the generation of the observed baryon asymmetry. In
Appendix~\ref{lepto} we provide a possible path to this direction
which shows that our setup preserves many of the successful
features of the post-inflationary evolution extensively explored
in the context of \fhi\ with other \Kas\ -- see, e.g.,
Refs.~\cite{sm1,smii,zubair,smsu5b}. On the other hand, this stage
of the cosmological history may impose additional constraints on
the viable parameter space and helps identify the most compelling
version of \fhi.


\appendix

\newcommand{\mfi}{\ensuremath{m_{\rm FI}}}
\newcommand{\qh}{\ensuremath{q_h}}

\section{Stabilization of the Modulus Before \fhi}\label{app}

We here present a sample stabilization of $h$ working along the
ideas of \cref{hpana1, hpana2}. Namely, we assume that $h$
transforms non-trivially under a (gauged) anomalous $U(1)_{\rm
FI}$ symmetry carrying charge $q_h$. Therefore, the function
$f(h)$ in \Eref{kh} must be a linear function of $\ln h$. In
particular, we assume
\beq f(h)=\ln(h/\mP)~~\mbox{with}~~h>\mP. \label{fdef}\eeq
Therefore $\Z$ and $\K$ in \Eref{kh} take the form
\beq
\Z=\ln|h/\mP|^2~~\mbox{and}~~\K=\bt\mP^2\ln\lf\ln|h/\mP|^2\rg.
\label{zkln}\eeq
The contribution to the D-term SUGRA potential due to $U(1)_{\rm
FI}$ is written as \cite{hpana2, hpana1}
\beq V_{\rm FI}= g^2{\rm D}^2_{\rm FI}/2~~~\mbox{with}~~~ {\rm
D}_{\rm FI}=\qh K_h h+ m_{\rm FI}^2, \label{vfi}\eeq
where we add a Fayet-Iliopoulos term $\mfi^2$ which naturally
assumes values of order $\mstr$ with contributions from the
non-vanishing $U(1)_{\rm FI}$ charges of the various fields of the
complete theory. Substituting $K$ in \Eref{ktot} into \Eref{vfi}
we obtain
\beq  V_{\rm FI}= g^2\mP^4 \lf\qh\bt\ln|h/\mP|^{-2}+\mfi^2/\mP^2
\rg^2/2,\label{vfih}\eeq
where we take into account that $S\ll h$ and $K_h\simeq\K_h$. A
global Minkowski vacuum occurs for
\beq
\mfi^2\vevi{\ln|h/\mP|^2}=-\qh\bt\mP^2~~~\Rightarrow~~~\vevi{h}=\mP\exp(-\qh\bt\mP^2/2\mfi^2).
\label{vevih}\eeq
It is easy to verify that $\vevi{V_{{\rm FI},h}}=0$ and
$\vevi{V_{{\rm FI},hh}}=0$. Imposing the normalization condition
of \Eref{vevh}, we find
\beq
-\frac{\qh\bt\mP^2}{2\mfi^2}=\frac12~~~\Rightarrow~~~\mfi=\sqrt{-\qh\bt}\mP.
\label{mfi}\eeq
Given that values $\bt=-1$ and $-2$ are encountered in our scheme,
we may achieve $\mfi\sim\mstr$ if $\qh\sim1/10$.

\section{GUT-Scale \fhi\ within \msgr} \label{msgr}

\renewcommand{\arraystretch}{1.2}
\begin{table}[!t]
\begin{center}
\begin{tabular}{|l|lll||lll|}
\hline
{\sc Cases}&\multicolumn{3}{|c||}{{\sc Type} I \fhi}&\multicolumn{3}{|c|}{{\sc Type} II \fhi}\\
\hline
&$p=2$&$p=3$&$p=4$&$q=3$&$q=5$&$q=7$\\ \hline\hline
$\sgx/10^{17}~\GeV$ & $3.165$&$4.57$&$4.965$&$1.556$&$3.11$&$3.683$\\
$\mst/\mstr$ & $0.85$&$0.17$&$0.11$&$5.4$&$0.09$&$0.05$\\
\hline
$\kp$&\multicolumn{3}{|c||}{$>\sqrt{4\pi}$}&$0.9$&\multicolumn{2}{|c|}{$>\sqrt{4\pi}$}\\
\hline
$M/10^{15}~\GeV$& $1.9$&$3.2$&$3.6$&$0.85$&$1.9$&$2.4$\\
$\sigma_{\rm
f}/10^{17}~\GeV$&$1.34$&$1.45$&$1.42$&$0.85$&$1.22$&$1.24$\\
$N_{\rm I*}$ & $49.3$&$49.6$&$49.7$&$48.7$&$49.1$&$49.7$\\\hline
$n_{\rm s}$ & $1.04$&$1.1$&$1.12$&$0.987$&$1.03$&$1.06$\\
$-\alpha_{\rm s}/10^{-3}$ & $1.65$&$4.5$&$5.9$&$0.87$&$16$&$2.4$\\
$r/10^{-5}$ & $1.2$&$9.6$&$1.5$&$0.048$&$1.2$&$3$\\\hline
$\msn/10^{15}~\GeV$ &
$0.37$&$1.5$&$2.6$&$0.07$&$0.59$&$1.3$\\\hline
\end{tabular}
\end{center}
\caption[]{\sl\small Input and output parameters compatible with
Eqs.~(\ref{nhi}), (\ref{prob}) and (\ref{Mgut}) within mSUGRA for
\fhia\ or \fhib\ and selected $p$ or $q$
respectively.}\label{tabm}
\end{table}

\renewcommand{\arraystretch}{1.}

To appreciate the improvements offered by our versions of \fhi\
regarding the inflationary predictions, we compare them with the
corresponding ones within mSUGRA \cite{hinova,okada}. In the
latter case, the inflationary  part of \Kap, $K_{\rm I}$ in
\Eref{ktot} is given by
\beq K_{\rm I}=|S|^2\label{kmsgr} \eeq
and the coefficients in expansion of \Eref{vhi} read \cite{hinova}
\beq \ck=\ckx=\ckh=0~~\mbox{and}~~\ckf=1/2. \label{cikmsgr}\eeq
We remark that the $\eta$ problem is spontaneously solved since we
obtain $\ck=0$ by construction. Our proposals are directly
comparable with \msgr\ since the same relation is achieved by
imposing the conditions in \Eref{shbt} or (\ref{nc}) within
\shsgr\ or \nsgr, respectively. On the other hand, $\ckf$ turns
out to be rather high in \msgr. Note that $\ckf=0$ within \shsgr\
and $0.046$ within \nsgr\ for the inputs of \Fref{fig3}.

Our results for \msgr\ are presented in \Tref{tabm} for \fhia\ or
\fhib\ and the same $p$ or $q$ values adopted in \Tref{tab} and
\Fref{fig1} and \ref{fig2}. We remark that $\sgx$ and $M$ slightly
increase relatively to their values in \shsgr\ whereas $|\as|, r$
and $\msn$ result one order of magnitude larger. These
ramifications shift upward $\ns$ which lies well above the
observational upper bounds of \eqs{nsact}{nsspt}. Only in the case
with $q=3$, where the simplified version of \fhi\ in \cref{simple,
simple5} is possible, $\ns$ turns out to be close to the upper
bound in \Eref{nsact}. Let us emphasize, once more, that the
constraint of \Eref{Mgut} plays a crucially role in this analysis.
Indeed, relaxing this requirement, we may achieve predictions
closer to the SUSY results discussed in Sec.~\ref{res1}. This
limitation of mSUGRA has also been highlighted in
Ref.~\cite{okada}. As a bottom line, GUT-scale \fhi\ within \msgr\
can be observationally excluded. \\

\newcommand{\rhn}{\ensuremath{N^c}}
\newcommand{\dphc}{\ensuremath{\delta\Phi}}
\newcommand{\dphcb}{\ensuremath{\delta\bar\Phi}}
\newcommand{\dphp}{\ensuremath{\delta\Phi_+}}
\newcommand{\ws}{\ensuremath{w_S}}
\newcommand{\wh}{\ensuremath{w_{\dphp}}}
\newcommand{\rhna}{\ensuremath{N^c_1}}
\newcommand{\rhni}{\ensuremath{N^c_i}}
\newcommand{\mrh[1]}{\ensuremath{M_{#1N^c}}}
\newcommand{\lrh[1]}{\ensuremath{\lambda_{#1N^c}}}
\newcommand{\Gr}{\ensuremath{\widetilde{G}}}
\newcommand{\Gm[1]}{\ensuremath{\Gamma_{#1}}}
\newcommand{\Gmi}{\ensuremath{\Gamma_{\rm I}}}
\def\ve{\varepsilon}

\section{Reheating Process, Lepton-Asymmetry and Gravitino
Abundances} \label{lepto}

We exemplify here a possible post-inflationary completion of our
models which clarifies the fact that $S$ oscillations dominate
over those of the Higgs fields ($\phcb-\phc$ or $\Psi$) and
verifies that a scenario of non-thermal leptogenesis consistent
with the gravitino ($\Gr$) constraint is implementable.

We focus on \fhia\ which allows for a direct coupling of $\phcb$
to right-handed neutrinos $\rhni$ via the superpotential term
\cite{sm1}
\beq  \lrh[i](\phcb \rhni)^2/\mstr\label{wrhn}\eeq
where $\phcb$ represents the SM singlet direction in $\phcb$. This
coupling is compatible with the symmetries of $W$ if
$2R(\rhni)=R(W)$ and $(B-L)(\phcb)=-(B-L)(\rhni)$. The coupling
above provides $\rhni$ with mass
\beq  \mrh[1]=2\lrh[1]\vev{\Phi}^2/\mstr\label{mrhn}\eeq
and allows for the decay into $\rhni$ of the inflaton system --
which consists of $S$ and $\dphp=(\dphcb+\dphc)/\sqrt{2}$, where
$\dphcb=\phcb-\vev{\phcb}$ and $\dphc=\phc-\vev{\phc}$ -- via the
common decay width
\beq
\Gmi=\frac{1}{32\pi}\gamma^2\msn~~\mbox{with}~~\gamma=\mrh[1]/\sqrt{2}\vev{\Phi}.\eeq
Here we assume that only the decay into the lightest $\rhni$ is
possible -- without competitive decay channels into the MSSM Higgs
superfields $H_u$ and $H_d$ as in \cref{zubair,phi} -- and
\beq\label{kin} 10\Trh\lesssim\mrh[1]\lesssim\msn/2.\eeq
The out-of-equilibrium decay of $\rhni$ generates an
$L$-asymmetry, which is not erased due to the left-most inequality
in \Eref{kin}, and obtains a maximal value \cite{sasa}
\beq\label{el} \ve_L = -\frac {3}{8\pi}\frac{m_{\nu_\tau}
\mrh[1]}{\vev{H_u}^2} \>\>\mbox{where}\>\>
m_{\nu_\tau}=\sqrt{\Delta m^2_\oplus}\simeq0.05\>{\rm eV} \eeq
is the mass of heaviest light neutrino $\nu_\tau$ and $\Delta
m^2_\oplus$ \cite{neutop} the atmospheric neutrino mass-squared
difference. Also, we set $\vev{H_u}=174~\GeV$ adopting the large
$\tan\beta$ regime of MSSM.

During the period of oscillations -- when $\msn\gg H_{\rm I0}$
with $H_{\rm I0}$ being defined in \Eref{m2sgbH} -- the energy
densities $\rho_S$ and $\rho_{\dphp}$ of the $S$ and $\dphp$
respectively, the energy density $\rho_{\rm R}$ of the produced
radiation, and the number densities $n_L$ of the leptons and
$n_{\Gr}$ of the $\Gr$'s satisfy the following Boltzmann equations
-- cf. Ref.~\cite{phi}:
\beqs\begin{eqnarray} && \dot \rho_S+3(1+\ws)H\rho_S+(1+\ws)\Gmi
\rho_S=0,\label{nf}\\
&& \dot\rho_{\dphp}+3(1+\wh)H\rho_{\dphp}+(1+\wh)\Gmi\rho_{\dphp}=0,\label{nfb} \\
&& \dot\rho_{\rm R}+4H\rho_{\rm R}-(1+\ws)\Gmi\rho_{S}
-(1+\wh)\Gmi\rho_{\dphp}=0,\label{rR}\\
&& \dot n_{L}+3Hn_{L}-2\ve_{L}\Gmi\lf(1+\ws) \rho_S+(1+\wh)\rho_{\dphp}\rg/\msn=0,\label{nL}\\
&& \dot n_{\Gr}+3Hn_{\Gr}-C_{\Gr} \lf n^{\rm eq}\rg^2=0.\label{ng}
\end{eqnarray}\eeqs
Here the overdot denotes derivation with respect to the cosmic
time $t$ and $H$ is the Hubble parameter during this period which
is given by
\begin{equation} \label{Hini}
H=\frac{1}{\sqrt{3}\mP} \left(\rho_S +\rho_{\dphp}+\rho_{\rm R}
\right)^{1/2}.
\end{equation}
The barotropic indices $\ws$ and $\wh$ can be approximately
estimated from \Eref{wrh} taking into account \Eref{vfo} and
setting $\sg$ and $\phi$ close to their v.e.vs in \Eref{vev}.
Namely, we find $n=2$ and so $\ws=0$ for $\varphi=\sg$ in
\Eref{wrh}. Similarly, for $\varphi=\phi$ \Eref{vfo} yields $n=4p$
and so $\wh=(2p-1)/(2p+1)$. Also $C_{\Gr}$ and $n^{\rm eq}$ are a
collision term for $\Gr$ production and the equilibrium number
density of each bosonic relativistic species. They have
respectively the form
\beq C_{\Gr}\simeq \frac{3\pi}{16\zeta(3)\mP^2}\sum_{i=1}^{3} c_i
g_i^2 \ln\left(\frac{k_{i}}{g_i}\right)~~\mbox{and}~~n^{\rm eq}=
{\zeta(3)\over\pi^2}T^3,\eeq
where in writing $C_{\Gr}$ we take the limit of massless MSSM
gauginos and of zero top trilinear coupling constant. We take
\cite{grv} $(c_i)=(21.336,35.33,29.4)$, $g_i$ are the gauge
coupling constants of the MSSM, and $(k_i)=(1.097,1.38,3.07)$.
Finally, the temperature $T$ and the entropy density ${\sf  s}$
are found from the relations
\begin{equation} \rho_{\rm R}=\frac{\pi^2}{30}g_{\rm rh*}
T^4~~\mbox{and}~~{\sf s}=\frac{2\pi^2}{45}g_{\rm rh*} T^3.
\label{rs}\end{equation}
The system of Eqs.~(\ref{nf})-(\ref{ng}) can be solved under the
following initial conditions
\beq\rho_S(0)=\rho_{\dphp}(0)=\Vhio/2~~\mbox{and}~~\rho_{\rm
R}(0)=n_{\Gr}(0)=n_{L}(0)=0,\label{init} \eeq
where we assumed that the inflationary energy density is equally
distributed between the oscillatory subsystems $S$ and $\dphp$.

\begin{figure}[!t]
\begin{minipage}{0.48\textwidth}
   \includegraphics[width=\linewidth]{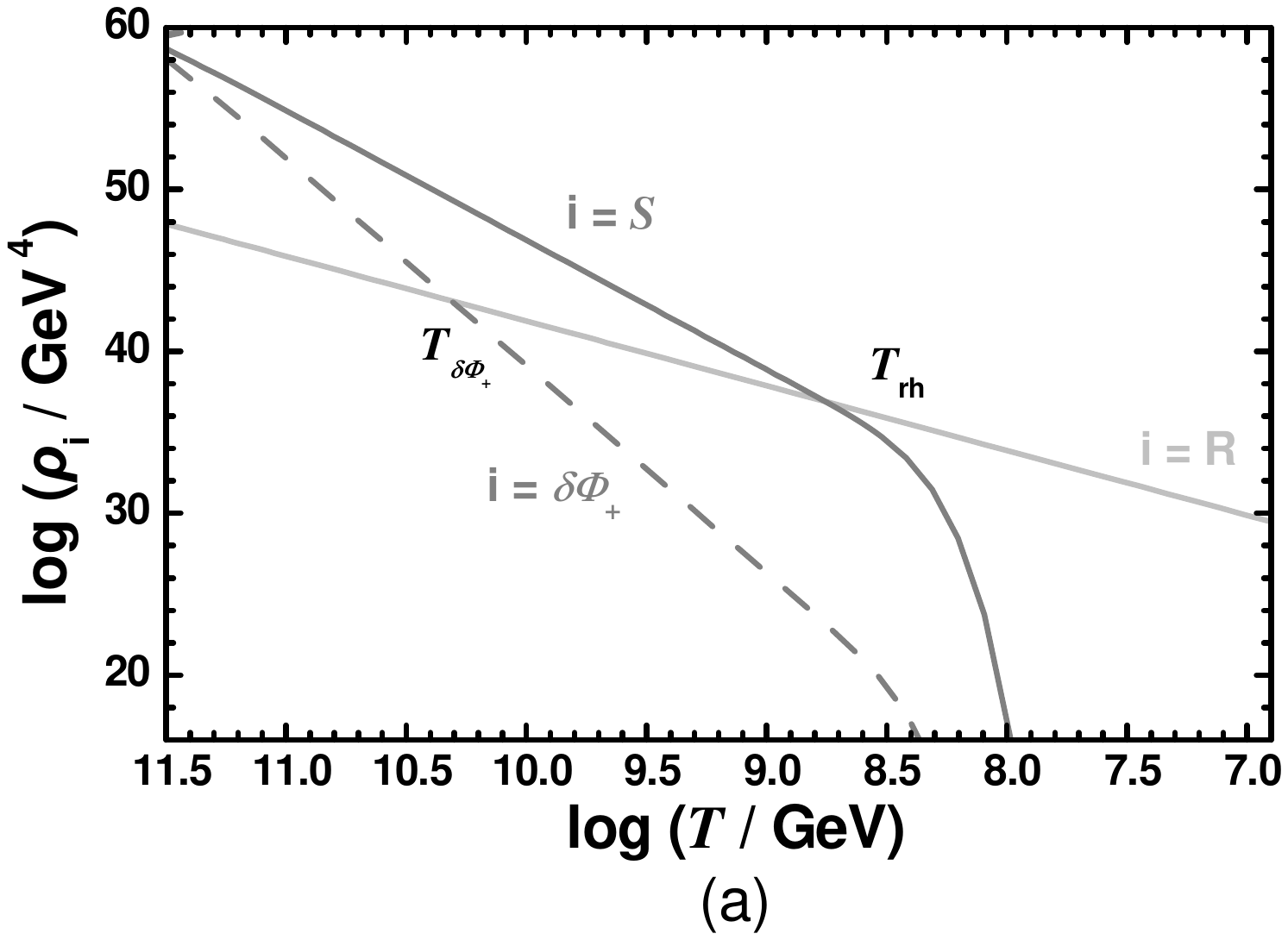}
\end{minipage}\hfill
\begin{minipage}{0.5\textwidth}
   \includegraphics[width=\linewidth]{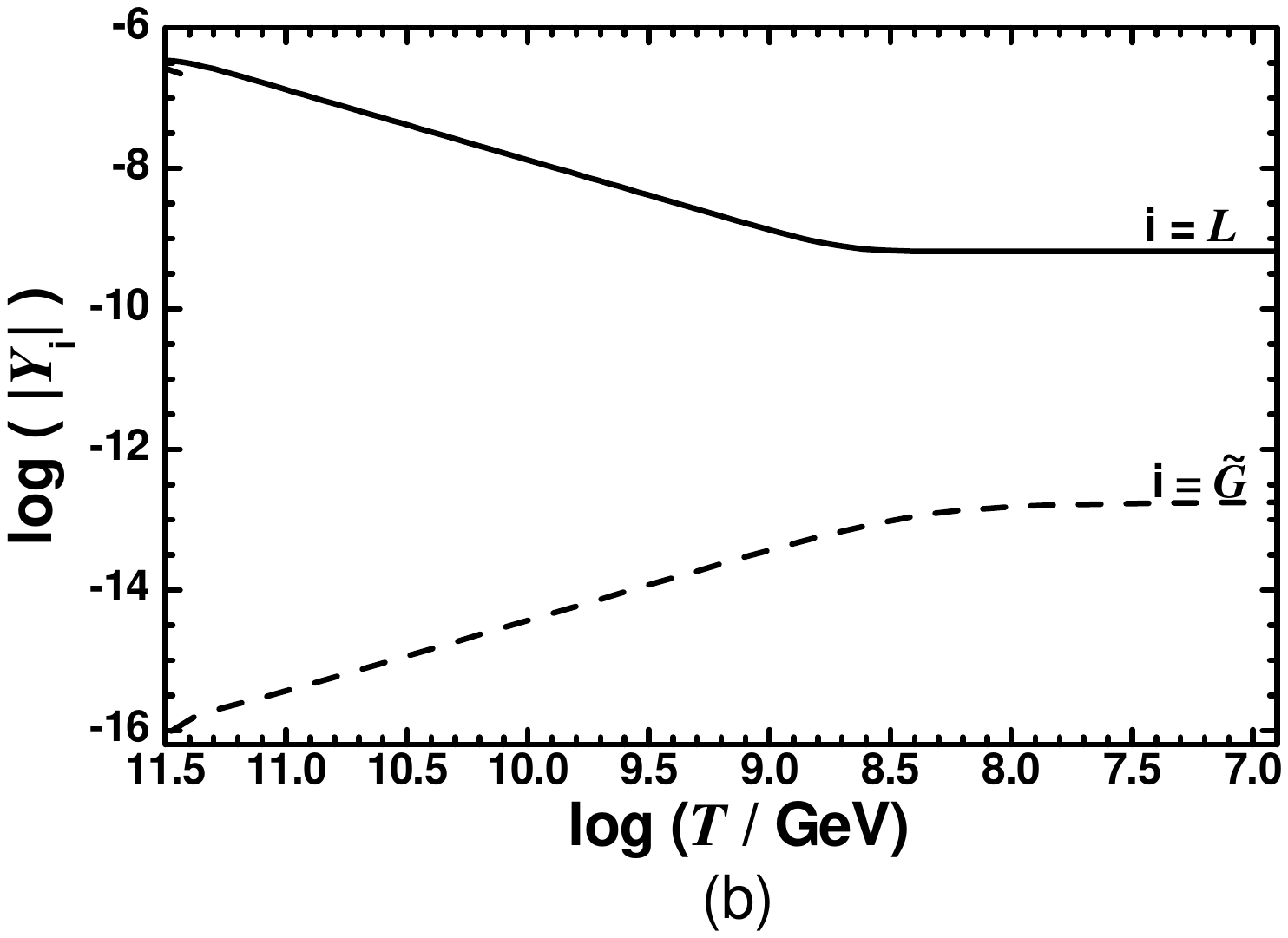}
\end{minipage}
\hfill \caption[]{\sl\small Evolution as functions of $\log T$ of
the quantities ({\sf\small a}) $\log\rho_i$ with $i=S$ (gray
line), $i=\dphp$ (gray dashed line) and $i={\rm R}$ (light gray
line), ({\sf\small b}) $\log |Y_L|$ (black solid line) and $\log
|Y_{\Gr}|$ (black dashed line). We take the values of the
parameters listed in \Tref{tab} with $p=2$ and
$\mrh[1]=10^{11}~\GeV$.}\label{fig5}
\end{figure}

In Fig.~\ref{fig5} we present the numerical solution of \Eref{nf}
-- (\ref{ng}) for the values of the parameters given in the
left-most column of \Tref{tab} (with $p=2$) and
$\mrh[1]=10^{11}~\GeV$. These parameters yield $H_{\rm
I0}\simeq2.6\cdot10^{11}~\GeV\ll \msn$ and therefore the phase of
the oscillations starts immediately after the end of \fhi. Also we
obtain $\lrh[1]=7.6\cdot10^{-6}$, $\Gmi=1.64~\GeV$, $\ws=0$ and
$\wh=3/5$ -- see \Eref{wrh}. More specifically, in \Fref{fig5}{a}
we illustrate the cosmological evolution of the quantities
$\log\rho_S$ (dotted gray line), $\log\rho_{\dphp}$ (dashed gray
line) and $\log\rho_{\rm R}$ (gray line) as functions of $\log T$
whereas in \Fref{fig5}{b} we depict $\log |Y_L|$ (black solid
line) and $\log |Y_{\Gr}|$ (black dashed line)  as functions of
$\log T$.

From \Fref{fig5}{a} we observe that \fhi\ is followed by the
matter-dominated era driven by the decaying $S$ since the decay of
$\dphp$ occurs very early at $T=T_{\dphp}\simeq2\cdot
10^{10}~\GeV$ -- this temperature corresponds to the intersection
of the $\rho_{\dphp}$ and $\rho_{\rm R}$ lines -- due to the
strong hierarchy between $\ws=0$ and $\wh=3/5$. The $S$
oscillations continue until $\rho_S$ meets $\rho_{\rm R}$ at
$\Trh=5.7\cdot10^{8}~\GeV$. This numerical result is in excellent
agreement with the estimate obtained by using $\wrh=\ws$ in the
approximate formula of $\Trh$ in \cref{phi}. After reheating, the
universe enters a conventional radiation dominated era. Therefore,
although our scenario involves two oscillatory systems, $\dphp$
and $S$, the final $\Trh$ can be accurately computed by
considering solely the system with the lowest $w$. Obviously, this
conclusion remains valid for larger $p$ values too.

In \Fref{fig5}{b} we depict the cosmological evolution of the
absolute values of the lepton-asymmetry and $\Gr$ abundances
defined as follows
\beq Y_{L}=n_{L}/{\sf s}~~\mbox{and}~~Y_{\Gr}=n_{{\Gr}}/{\sf
s}.\eeq
We see that $|Y_{L}|$ and $Y_{\Gr}$ reach their constant values
equal to $2.3\times10^{-10}$ and $1.8\times10^{-13}$ respectively
immediately after the completion of $S$ decay, close to
$T\simeq\Trh$ -- note that $Y_L$ originates almost exclusively
from the lepton asymmetry $2\ve_{L}$ generated by the $S$ decay.
$Y_L$ is partially converted through sphaleron effects into a
yield of the $B$ asymmetry of the universe $Y_B$ which meets its
observational value \cite{act} since
$-0.35Y_{L}(\Trh)=8.7\cdot10^{-11}$. Moreover, the present value
$Y_{\Gr}$ is compatible with the upper bound posed by the
requirements \cite{kohri} of Big Bang nucleosynthesis for
$m_{\Gr}\gtrsim10~\TeV$ -- cf. \cref{phi}. Therefore, non-thermal
leptogenesis is a realistic possibility within \fhia.

The case of \fhib\ requires further investigation since the
coupling in \Eref{wrhn} has to be modified and so the reheating
process, the generation of neutrino masses and the leptogenesis
mechanism depend on the structure of the complete model -- see,
e.g., \cref{smii, smsu5b}.

\newcommand\jcap[3]{{\it JCAP }{\bf #1}, #3 (#2)}
\renewcommand\jhep[3]{{\it JHEP }{\bf
#1}, #3 (#2)}
\newcommand{\arxiv}[1]{{\tt arXiv:#1}}
\def\prd#1#2#3{{\sl Phys. Rev. D }{\bf #1}, #3 (#2)}
\def\prdn#1#2#3#4{{\sl Phys. Rev. D }{\bf #1}, no~#4, #3 (#2)}



\begin{thebibliography}{100}

\bibitem{lect} G.~Lazarides, {\it Basics of inflationary cosmology}, {\sl J.
Phys. Conf. Ser.} \textbf{53}, 528 (2006) [{\tt hep-ph/0607032}].


\bibitem{hinova} R. Armillis and C. Pallis,
{\it Implementing Hilltop F-term Hybrid Inflation in
Supergravity,} in {\sl Recent Advances in Cosmology, edited by A.
Travena and B. Soren (Nova Science Publishers Inc., New York,
2013)}  \arxiv{1211.4011}.


\bibitem{sm} G. Lazarides and C. Panagiotakopoulos, {\it Smooth hybrid
inflation}, {\sl Phys. Rev. D} {\bf 52}, R559 (1995) [{\tt hep-
ph/9506325}].

\bibitem{ini} G.~Lazarides, C.~Panagiotakopoulos and N.D.~Vlachos, {\it Initial
conditions for smooth hybrid inflation}, {\sl Phys. Rev. D}
\textbf{54}, 1369 (1996) [{\tt hep-ph/9606297}].

\bibitem{sm1} R.~Jeannerot, S.~Khalil and G.~Lazarides, {\it Leptogenesis
in smooth hybrid inflation}, {\sl Phys. Lett. B} {\bf506}, 344
(2001) [{\tt hep-ph/0103229}].


\bibitem{nsm} G. Lazarides and A. Vamvasakis, {\it New smooth hybrid inflation},
{\sl Phys. Rev. D} {\bf 76}, 083507 (2007) [\arxiv{0705.3786}].




\bibitem{smy} M.~Yamaguchi and J.~Yokoyama,
{\it Smooth hybrid inflation in supergravity with a running
spectral index and early star formation,} {\it Phys. Rev. D }
\textbf{70}, 023513 (2004) [\hepph{0402282}].



\bibitem{smk} M.~Kawasaki, N.~Kitajima and K.~Nakayama,
{\it Smooth hybrid inflation in a supersymmetric axion model,}
\textit{Phys. Rev. D } \textbf{87}, no.3, 035010 (2013)
[\arxiv{1211.6516}].




\bibitem{smii} S.~Khalil, Q.~Shafi and A.~Sil, {\it Smooth Hybrid Inflation and
Non-Thermal Type II Leptogenesis}, {\sl Phys. Rev. D} \textbf{86},
073004 (2012) [\arxiv{1208.0731}].


\bibitem{smsu5b} W.~Ahmed, A.~Karozas, G.~K.~Leontaris and U.~Zubair, {\it Smooth
hybrid inflation with low reheat temperature and observable
gravity waves in $SU(5) \times U(1)_{\chi}$ super-GUT}, {\sl JCAP}
\textbf{06}, no. 06, 027 (2022) [\arxiv{2201.12789}].


\bibitem{senoguz} V.N.~Senoguz and Q.~Shafi, {\it Testing supersymmetric grand
unified models of inflation}, {\sl Phys. Lett. B} {\bf 567}, 79
(2003) [{\tt hep-ph/0305089}].



\bibitem{mur} M.~ur Rehman, V.N.~Senoguz and Q.~Shafi, {\it Supersymmetric And
Smooth Hybrid Inflation In The Light Of WMAP3}, {\sl Phys. Rev. D}
\textbf{75}, 043522 (2007) [{\tt hep-ph/0612023}].



\bibitem{simple} M.U.~Rehman and Q.~Shafi, {\it Simplified Smooth Inflation with
Observable Gravity Waves}, {\sl Phys. Rev. D } \textbf{86}, 027301
(2012) [\arxiv{1202.0011}].

\bibitem{simple5} M.U.~Rehman and
U.~Zubair, {\it Simplified Smooth Hybrid Inflation in
Supersymmetric SU(5),} {\sl Phys. Rev. D } \textbf{91}, 103523
(2015) [\arxiv{1412.7619}].


\bibitem{zubair} U.~Zubair, {\it Smooth {\ensuremath{\mu}}-hybrid and non-minimal
Higgs inflation in $SU(4)_{\rm C} {\times} SU(2)_{L} {\times}
SU(2)_{R}$ with observable gravitational waves,} {\sl JCAP}
\textbf{02}, 033 (2025) [\arxiv{2403.13991}].

\bibitem{okada} N.~Okada and O.~Seto, {\it Smooth hybrid inflation in light of ACT
DR6 data}, {\sl Phys. Rev. D} \textbf{112}, no.~8, 083549 (2025)
[\arxiv{2506.15965}].











\bibitem{pseudo1} S.~Antusch, D.~Nolde and M.U.~Rehman, {\it Pseudosmooth Tribrid
Inflation,} {\sl JCAP} \textbf{08}, 004 (2012)
[\arxiv{1205.0809}].

\bibitem{pseudo2} M.A.~Masoud, M.U.~Rehman and Q.~Shafi,
{\it Pseudosmooth Tribrid Inflation in $SU(5)$,} {\sl JCAP}
\textbf{04}, 041 (2020) [\arxiv{1910.07554}].



\bibitem{fhi} G.R.~Dvali, Q.~Shafi and R.K.~Schaefer, {\it Large scale structure
and supersymmetric inflation without fine tuning}, {\sl Phys. Rev.
Lett} \textbf{73}, 1886 (1994) [{\tt hep-ph/9406319}].


\bibitem{sh} R. Jeannerot  {\it et al.}, {\sl Inflation and monopoles in
supersymmetric SU(4)C x SU(2)(L) x SU(2)(R)}, \jhep{10}{2000}{012}
[{\tt hep-ph/0002151}].


\bibitem{nsh} R. Jeannerot, S. Khalil and G. Lazarides, {\it New shifted hybrid
inflation}, {\sl J. High Energy Phys.} {\bf 07}, 069 (2002)
[\hepph{0207244}].


\bibitem{ssh} G. Lazarides, I.N.R. Peddie and A. Vamvasakis, {\it Semi-shifted
hybrid inflation with B-L cosmic strings}, {\sl Phys. Rev. D} {\bf
78}, 043518 (2008) [\arxiv{0804.3661}].


\bibitem{su5sh} S.~Khalil, M.U.~Rehman, Q.~Shafi and E.A.~Zaakouk, {\it Inflation
in Supersymmetric SU(5)}, {\sl Phys. Rev. D} \textbf{83}, 063522
(2011) [\arxiv{1010.3657}].

\bibitem{mush} G.~Lazarides, M.U.~Rehman, Q.~Shafi and F.K.~Vardag,
{\it Shifted $\mu$-hybrid inflation, gravitino dark matter, and
observable gravity waves,} {\sl Phys. Rev. D }\textbf{103}, no.3,
035033 (2021) [\arxiv{2007.01474}].

\bibitem{su4sh} A.~Afzal, M.~Mehmood, M.~U.~Rehman and Q.~Shafi, {\it
Supersymmetric hybrid inflation and current-carrying metastable
cosmic strings in $SU(4)_c \times SU(2)_L \times U(1)_R$},
\arxiv{2308.11410}.


\bibitem{plin} Y.~Akrami {\it et al.} [\plk\ Collaboration], {\it Planck
2018 results. X. Constraints on inflation},  {\sl Astron.
Astrophys}. \textbf{641}, A10 (2020) [\arxiv{1807.06211}].



\bibitem{bcp} P.A.R. Ade \etal\ [BICEP, Keck collaboration], {\it Improved
Constraints on Primordial Gravitational Waves using Planck, WMAP,
and BICEP/Keck Observations through the 2018 Observing Season},
 {\sl Phys. Rev. Lett} {\bf 127}, 151301  (2021) [\arxiv{2110.00483}].


\bibitem{act} T.~Louis \textit{et al.} [ACT Collaboration],
{\it The Atacama Cosmology Telescope: DR6 Power Spectra,
Likelihoods and $\Lambda$CDM Parameters}, \arxiv{2503.14452}.

\bibitem{actin} E.~Calabrese \textit{et al.} [ACT  Collaboration],
{\it The Atacama Cosmology Telescope: DR6 Constraints on Extended
Cosmological Models}, \arxiv{2503.14454}.


\bibitem{desi} A.G. Adame \etal\ [DESI collaboration], {\it DESI 2024 VI: cosmological constraints from
the measurements of baryon acoustic oscillations}, {\sl JCAP}
\textbf{02}, 021 (2025) [\arxiv{2404.03002}].

\bibitem{spt} E. Camphuis \etal\ [SPT-3G Collaboration], {\it SPT-3G D1:
CMB temperature and polarization power spectra and cosmology from
2019 and 2020 observations of the SPT-3G Main field},
\arxiv{2506.20707}.







\bibitem{fhi1}  M.U.~Rehman and Q.~Shafi,
{\it Supersymmetric Hybrid Inflation in light of Atacama Cosmology
Telescope Data Release 6, Planck 2018 and LB-BK18}, {\sl Phys.
Rev. D } \textbf{112}, no.~2, 023529 (2025) [\arxiv{2504.14831}].



\bibitem{fhi2} C. Pallis, {\it F-Term Hybrid Inflation, Metastable Cosmic Strings and
Low Reheating in View of ACT,} {\sl PoS CORFU} \textbf{2024}, 206
(2025) [\arxiv{2504.20273}].

\bibitem{fhi3}
M.N.~Ahmad and M.U.~Rehman, {\it Supersymmetric hybrid inflation
with K{\"a}hler-induced R-symmetry breaking}, {\sl JCAP }
\textbf{08}, 061 (2025) [\arxiv{2506.23244}].

\bibitem{fhi4} A.~Moursy and Q.~Shafi,
{\it Waterfall phase in supersymmetric hybrid inflation,} {\sl
JHEP} \textbf{01}, 162 (2026) [\arxiv{2507.10460}].

\bibitem{fhi5} N.~Okada and Q.~Shafi,
{\it Split supersymmetry and hybrid inflation in light of Atacama
Cosmology Telescope DR6 data,} \arxiv{2507.16246}.




\bibitem{act0} R.~Kallosh, A.~Linde and D.~Roest,
{\it ACT, SPT, and chaotic inflation,} {\sl Phys. Rev. Lett.}
\textbf{135}, no.16, 161001 (2025) [\arxiv{2503.21030}].

\bibitem{act2} Z.~Yi, X.~Wang, Q.~Gao and Y.~Gong, {\it Potential
Reconstruction from ACT Observations Leading to Polynomial
$\alpha$-Attractor,} \arxiv{2505.10268}.


\bibitem{act4}  M.~He, M.~Hong and K.~Mukaida,
{\it Increase of $\ns$ in regularized pole inflation \&
Einstein-Cartan gravity,} \arxiv{2504.16069}.



\bibitem{gup} H.~Heidarian, M.~Solbi, S.~Heydari and K.~Karami, {\it
$\alpha$-attractor inflation modified by GUP in light of ACT
observations,}  {\sl Phys. Lett. B} {\bf 869}, 139833 (2025)
[\arxiv{2506.10547}].

\bibitem{oxf} W.J.~Wolf,
{\it Inflationary attractors and radiative corrections in light of
ACT,} {\sl JCAP} \textbf{02}, 088 (2026) [\arxiv{2506.12436}].


\bibitem{indi} S.~Choudhury, B.~Gulnur, S.K.~Singh and K.~Yerzanov,
{\it What new physics can we extract from inflation using the ACT
DR6 and DESI DR2 Observations?,} \arxiv{2506.15407}.

\bibitem{kina} Q.~Gao, Y.~Qian, Y.~Gong and Z.~Yi,
{\it Observational constraints on inflationary models with
non-minimally derivative coupling by ACT,} \arxiv{2506.18456}.

\bibitem{actlee} J.~Han, H.M.~Lee and J.H.~Song,
{\it Higgs pole inflation with loop corrections in light of ACT
results,} \arxiv{2506.21189}.



\bibitem{rhc} R.~Mondal, S.~Mondal and A.~Chakraborty, {\it Constraining
Reheating Temperature, Inflaton-SM Coupling and Dark Matter Mass
in Light of ACT DR6 Observations,} \arxiv{2505.13387}.

\bibitem{rhb} L. Liu, Z. Yi, and Y. Gong, {\it Reconciling Higgs Inflation with ACT
Observations through Reheating}, \arxiv{2505.02407}.

\bibitem{rha} S.~Maity,
{\it ACT-ing on inflation: Implications of non Bunch-Davies
initial condition and reheating on single-field slow-roll models,}
\arxiv{2505.10534}.

\bibitem{act5} M.R.~Haque, S.~Pal and D.~Paul, {\it ACT DR6 Insights on the
Inflationary Attractor models and Reheating,} \arxiv{2505.01517}.

\bibitem{act7}  E.G.M.~Ferreira, E.~McDonough, L.~Balkenhol, R.~Kallosh, L.~Knox
and A.~Linde, {\it The BAO-CMB Tension and Implications for
Inflation,} \arxiv{2507.1245}.

\bibitem{nmact}  Q. Gao, Y. Gong, Z. Yi and F. Zhang, {\it Non-minimal coupling in
light of ACT,} \arxiv{2504.15218}.

%
\bibitem{maity} M.R.~Haque and D.~Maity, {\it Minimal Plateau Inflation in light
of ACT DR6 Observations,} {\sl Phys. Lett. B} \textbf{873}, 140187
(2026) [\arxiv{2505.18267}].

\bibitem{reh8} L.Y.~Chen, R.~Zha and F.Y.~Zhang, {\sl Probing Reheating in a
Decaying Oscillatory Inflationary Model with Latest ACT
Constraints,} \arxiv{2508.16538}.


\bibitem{actattr} C. Dioguardi, A.J. Iovino and A. Racioppi, {\it Fractional
attractors in light of the latest ACT observations}, {\sl Phys.
Lett. B} \textbf{868}, 139664 (2025) [\arxiv{2504.02809}].

\bibitem{act1} J.~McDonald, {\it Higgs Inflation with Vector-Like Quark
Stabilisation and the ACT spectral index,} \arxiv{2505.07488}.

\bibitem{actj} J.~McDonald, {\it Unitarity-Conserving Non-Minimally Coupled Inflation and the
ACT Spectral Index,} {\sl Phys. Rev. D} \textbf{112}, no.12,
123525 (2025) [\arxiv{2506.12916}].

\bibitem{yin}  W.~Yin, {\it Higgs-like inflation under ACTivated mass,}
{\sl JCAP} \textbf{09}, 062 (2025) [\arxiv{2505.03004}].

\bibitem{actpal} C.~Pallis, \textit{Kinetically Modified Palatini
Inflation Meets ACT Data}, {\sl Phys. Lett. B }{\bf 868}, 139739
(2025) [\arxiv{2505.23243}].

\bibitem{act3} Z.Z.~Peng, Z.C.~Chen and L.~Liu, {\it The polynomial potential
inflation in light of ACT observations,} \arxiv{2505.12816}.


\bibitem{act8} A.~Mohammadi, Yogesh and A.~Wang,
{\it Power Law Plateau Inflation and Primary Gravitational Waves
in the light of ACT,} \arxiv{2507.06544}.

\bibitem{act6} I.D.~Gialamas, A.~Karam, A.~Racioppi and M.~Raidal,
{\it Has ACT measured radiative corrections to the tree-level
Higgs-like inflation?,} {\sl Phys. Rev. D} \textbf{112}, no.10,
103544 (2025) [\arxiv{2504.06002}].

\bibitem{actellis}
J.~Ellis, M.A.G.~Garc{\'\i}a, N.~Nagata, D.V.~Nanopoulos and
K.A.~Olive, {\it Deformations of Starobinsky Inflation in No-Scale
SU(5) and SO(10) GUTs,} {\sl JCAP} \textbf{12}, 038 (2025)
[\arxiv{2508.13279}].

\bibitem{acttamv} I.D. Gialamas, T. Katsoulas and K. Tamvakis, {\it Keeping the
relation between the Starobinsky model and no-scale supergravity
ACTive}, {\sl JCAP} \textbf{09}, 060 (2025) [\arxiv{2505.03608}].

\bibitem{ketov} A. Addazi,  Y. Aldabergenov and S.V. Ketov, {\it Curvature
corrections to Starobinsky inflation can explain the ACT results},
{\sl Phys. Lett. B} \textbf{869}, 139883 (2025)
[\arxiv{2505.10305}].

\bibitem{r2a} Yogesh, A. Mohammadi, Q. Wu and
T. Zhu, {\it Starobinsky like inflation and EGB gravity in the
light of ACT}, {\sl JCAP} \textbf{10}, 010 (2025)
[\arxiv{2505.05363}].

\bibitem{r2b} M.R. Haque, S. Pal and D. Paul, {\it Improved Predictions on
Higgs-Starobinsky Inflation and Reheating with ACT DR6 and
Primordial Gravitational Waves}, \arxiv{2505.04615}.

\bibitem{r2drees} M. Drees and Y. Xu, {\it Refined Predictions for Starobinsky Inflation
and Post-inflationary Constraints in Light of ACT}, {\sl Phys.
Lett. B }\textbf{867}, 139612 (2025) [\arxiv{2504.20757}].

\bibitem{r2mans} W.~Ahmed and M.U.~Rehman, {\it Radiatively Corrected Starobinsky
Inflation and Primordial Gravitational Waves in Light of ACT
Observations,} \prdn{112}{2025}{063519}{6} \arxiv{2506.18077}.

\bibitem{r2li} J.~Kim, X.~Wang, Y.l.~Zhang and Z.~Ren,
{\it Enhancement of primordial curvature perturbations in
$R^3$-corrected Starobinsky-Higgs inflation,} \arxiv{2504.12035}.

\bibitem{heavy} S.~Aoki, H.~Otsuka and R.~Yanagita,
{\it Heavy Field Effects on Inflationary Models in Light of ACT
Data,} {\sl JCAP} \textbf{11}, 088 (2025) [\arxiv{2509.06739}].

\bibitem{actpole} C.~Pallis,
{\it ACT-Inspired K\"ahler-Based Inflationary Attractors,}
\jcap{09}{2025}{061} [\arxiv{2507. 02219}].

\bibitem{phi} C.~Pallis, {\sl Updating GUT-Scale Pole Higgs Inflation After ACT DR6,}
\prdn{113}{2026}{015033}{1} [\arxiv{2510.02083}].


\bibitem{das} S.~Das,
{\it Resurrecting vanilla power law inflation with the aid of
continuous spontaneous localization in the ACT era,} {\sl Phys.
Rev. D} \textbf{112}, no.~2, 023543 (2025) [\arxiv{2508.14602}].

\bibitem{salvio} A.~Salvio,
{\it Independent connection in action during inflation,} {\sl
Phys. Rev. D}  \textbf{112}, no.~6, L061301 (2025)
[\arxiv{2504.10488}].

\bibitem{aoki} S.~Aoki, H.~Otsuka and R.~Yanagita,
{\it Higgs-modular inflation,} {\sl Phys. Rev. D}
 \textbf{112}, no.4, 043505 (2025) [\arxiv{2504.01622}].

\bibitem{hai} M.~Hai, A.~R.~Kamal, N.~F.~Shamma and M.S.J.~Shuvo,
{\it Perturbative K{\"a}hler Moduli Inflation},
\arxiv{2506.08083}.




\bibitem{lofti} L.~Boubekeur and D.H.~Lyth, {\it Hilltop inflation}, {\sl JCAP}
\textbf{07}, 010 (2005) [{\tt hep-ph/0502047}].

\bibitem{fhim} G.~Lazarides and C.~Pallis, {\it Reducing the spectral index in F-term hybrid
inflation through a complementary modular inflation,} {\sl Phys.
Lett. B} \textbf{651}, 216 (2007) [\hepph{0702260}].


\bibitem{ibanez} L.E.~Iba\~nez and D.~Lust, {\it Duality anomaly cancellation,
minimal string unification and the effective low-energy Lagrangian
of 4-D strings}, {\sl Nucl. Phys. B} {\bf 382}, 305 (1992) [{\tt
hep-th/9202046}].

\bibitem{lust} D.~Lust, S.~Reffert and S.~Stieberger, {\it MSSM with soft SUSY
breaking terms from D7-branes with fluxes}, {\sl Nucl. Phys. B}
\textbf{727}, 264 (2005) [{\tt hep-th/0410074}].


\bibitem{eno7} J.~Ellis, D.~Nanopoulos and
K.~Olive, {\it Starobinsky-like Inflationary Models as Avatars of
No-Scale Supergravity}, \jcap{10}{2013}{009} [\arxiv{1307.3537}].


\bibitem{kelar} C.~Pallis, {\it K{\"a}hler Potentials for Hilltop F-Term Hybrid
Inflation}, {\sl JCAP } \textbf{04}, 024 (2009) [\arxiv{0902.
0334}].


\bibitem{antu} S.~Antusch, M.~Bastero-Gil, K.~Dutta, S.F.~King and P.M.~Kostka,
{\it Solving the eta-Problem in Hybrid Inflation with Heisenberg
Symmetry and Stabilized Modulus,} {\sl JCAP } \textbf{01}, 040
(2009) [\arxiv{0808.2425}].

\bibitem{asfhi} G.~Lazarides and C.~Pallis,
{\it Probing the supersymmetry-mass scale with F-term hybrid
inflation,} {\sl Phys. Rev. D} \textbf{108}, no.~9, 095055 (2023)
[\arxiv{2309.04848}].

\bibitem{hpana1}
C.~Panagiotakopoulos, {\it Hybrid inflation in supergravity with
$(SU(1, 1)/ U(1))^m$ Kahler manifolds}, {\sl Phys. Lett. B}
\textbf{459}, 473 (1999)  [{\tt hep-ph/9904284}].

\bibitem{hpana2} C.~Panagiotakopoulos, {\it Realizations of hybrid inflation in
supergravity with natural initial conditions}, {\sl Phys. Rev. D}
\textbf{71}, 063516 (2005) [{\tt hep-ph/0411143}].

\bibitem{eta} E.D.~Stewart, {\it Inflation, supergravity and superstrings}, {\sl
Phys. Rev. D} \textbf{51}, 6847 (1995) [{\tt hep-ph/ 9405389}].


\bibitem{shiftk0} M. Kawasaki, M. Yamaguchi and T. Yanagida,
{\it Natural chaotic inflation in supergravity}, {\sl Phys. Rev.
Lett.} {\bf 85}, 3572 (2000) [\hepph{0004243}].

\bibitem{shiftk} C.~Pallis and Q.~Shafi,
{\it From Hybrid to Quadratic Inflation With High-Scale
Supersymmetry Breaking,} {\sl Phys. Lett. B} \textbf{736}, 261
(2014) [\arxiv{1405.7645}].

\bibitem{alinde}  R.~Kallosh, A.~Linde, and D.~Roest,
{\it Superconformal Inflationary $a$-Attractors,}
\jhep{11}{2013}{198} [\arxiv{1311.0472}].


\bibitem{bernal} N.~Bernal, F.~Elahi, C.~Maldonado and J.~Unwin,
{\it Ultraviolet Freeze-in and Non-Standard Cosmologies,} {\sl
JCAP} \textbf{11}, 026 (2019) [\arxiv{1909.07992}].

\bibitem{lept}  G. Lazarides and Q. Shafi, {\it Origin of matter in inflationary cosmology},
\plb{258}{1991}{305}.

\bibitem{grv}
H.~Eberl, I.D.~Gialamas and V.C.~Spanos, {\it Gravitino thermal
production, dark matter, and reheating of the Universe,} {\sl
JCAP} \textbf{01}, 079 (2025) [\arxiv{2408.16043}].


\bibitem{sasa} S. Davidson and A. Ibarra, {\it A Lower bound on the right-handed
neutrino mass from leptogenesis}, {\sl Phys. Lett. B} {\bf 535},
25 (2002) [\hepph{0202239}].

\bibitem{neutop} I.~Esteban, M.C.~Gonzalez-Garcia, M.~Maltoni, I.~Martinez-Soler,
J.P.~Pinheiro and T.~Schwetz, {\it NuFit-6.0: updated global
analysis of three-flavor neutrino oscillations,}
\jhep{12}{2024}{216} [\arxiv{2410.05380}].

\bibitem{kohri}  M.~Kawasaki, K.~Kohri, T.~Moroi and Y.~Takaesu,
{\it Revisiting Big-Bang Nucleosynthesis Constraints on Long-Lived
Decaying Particles,} {\sl Phys. Rev. D }\textbf{97}, no.~2, 023502
(2018) [\arxiv{1709.01211}].


\end{thebibliography}
\end{document}